\documentstyle[11pt,aaspp4,flushrt]{article}

\slugcomment{\it To apper in The Astronomical Journal (May 2002)} 
\lefthead{Stern et al.}
\righthead{{\it Chandra} Observations of Lynx}

\def\ie{{i.e.,~}}
\def\eg{{e.g.,~}}
\def\etal{{et al.~}}

\def\deg{\ifmmode {^{\circ}}\else {$^\circ$}\fi}

\def\secper{\ifmmode \rlap.{^{s}}\else $\rlap{.}{^{s}} $\fi}

\def\kms{\ifmmode {\rm\,km\,s^{-1}}\else
    ${\rm\,km\,s^{-1}}$\fi}
\def\kmsMpc{\ifmmode {\rm\,km\,s^{-1}\,Mpc^{-1}}\else
    ${\rm\,km\,s^{-1}\,Mpc^{-1}}$\fi}
\def\ergAcm2{\ifmmode {\rm\,ergs\,cm^{-2}\,{\rm \AA}^{-1}}\else
    ${\rm\,ergs\,cm^{-2}\,\AA^{-1}}$\fi}
\def\cm2{\ifmmode {\rm\,cm^{-2}}\else
    ${\rm\,cm^{-2}}$\fi}
\def\ergcm2s{\ifmmode {\rm\,ergs\,cm^{-2}\,s^{-1}}\else
    ${\rm\,ergs\,cm^{-2}\,s^{-1}}$\fi}
\def\cgsdeg2{\ifmmode {\rm\,ergs\,cm^{-2}\,s^{-1}\,deg^{-2}}\else
    ${\rm\,ergs\,cm^{-2}\,s^{-1}\,deg^{-2}}$\fi}
\def\ergsHz{\ifmmode {\rm\,ergs\,s^{-1}\,Hz^{-1}}\else
    ${\rm\,ergs\,s^{-1}\,Hz^{-1}}$\fi}
\def\ergs{\ifmmode {\rm\,ergs\,s^{-1}}\else
    ${\rm\,ergs\,s^{-1}}$\fi}
\def\ergsA{\ifmmode {\rm\,ergs\,s^{-1}\,\AA^{-1}}\else
    ${\rm\,ergs\,s^{-1}\,\AA^{-1}}$\fi}
\def\WHz{\ifmmode {\rm\,W\,Hz^{-1}}\else
    ${\rm\,W\,Hz^{-1}}$\fi}
\def\WHzsr{\ifmmode {\rm\,W\,Hz^{-1}\,sr^{-1}}\else
    ${\rm\,W\,Hz^{-1}\,sr^{-1}}$\fi}
\def\ergscm2Hz{\ifmmode {\rm\,ergs\,cm^{-2}\,s^{-1}\,Hz^{-1}}\else
    ${\rm\,ergs\,cm^{-2}\,s^{-1}\,Hz^{-1}}$\fi}

\def\spose#1{\hbox to 0pt{#1\hss}}
\def\simlt{\mathrel{\spose{\lower 3pt\hbox{$\mathchar"218$}}
     \raise 2.0pt\hbox{$\mathchar"13C$}}}
\def\simgt{\mathrel{\spose{\lower 3pt\hbox{$\mathchar"218$}}
     \raise 2.0pt\hbox{$\mathchar"13E$}}}

\def\ciii{\ion{C}{3}] $\lambda$1909}
\def\cii{\ion{C}{2}] $\lambda$2326}
\def\neiv{[\ion{Ne}{4}] $\lambda$2424}
\def\mgii{\ion{Mg}{2} $\lambda$2800}
\def\nev{[\ion{Ne}{5}] $\lambda \lambda$3346, 3426}
\def\oii{[\ion{O}{2}] $\lambda$3727}

\def\neiii{[\ion{Ne}{3}] $\lambda$3869}
\def\hbeta{H$\beta$}
\def\oiii{[\ion{O}{3}] $\lambda$5007}

\def\halpha{H$\alpha$}

\def\plotfiddle#1#2#3#4#5#6#7{\centering \leavevmode
\vbox to#2{\rule{0pt}{#2}}
\includegraphics{#1}}

\begin{document}

\title{SPICES II.  Optical and Near-Infrared Identifications \\ 
of Faint X-Ray Sources from Deep {\it Chandra} Observations of Lynx$^1$}

\author{Daniel Stern\altaffilmark{2},
Paolo Tozzi\altaffilmark{3,4},
S.~A.~Stanford\altaffilmark{5,6},
Piero Rosati\altaffilmark{3,7},
Brad Holden\altaffilmark{5,6}, \\
Peter Eisenhardt\altaffilmark{2},
Richard Elston\altaffilmark{8},
K.~L.~Wu\altaffilmark{8},
Andrew Connolly\altaffilmark{9}, \\
Hyron Spinrad\altaffilmark{10},
Steve Dawson\altaffilmark{10},
Arjun Dey\altaffilmark{11}, 
\& Frederic H. Chaffee\altaffilmark{12}}

\altaffiltext{1}{Some of the data presented herein were obtained at the
W.M. Keck Observatory, which is operated as a scientific partnership
among the California Institute of Technology, the University of
California and the National Aeronautics and Space Administration. The
Observatory was made possible by the generous financial support of the
W.M. Keck Foundation.  Based on observations with the NASA/ESA {\it
Hubble Space Telescope}, obtained at the Space Telescope Science
Institute, which is operated by the Association of Universities for
Research in Astronomy, Inc. under NASA contract No. NAS5-26555.  Based
on observations at the Kitt Peak National Observatory, National Optical
Astronomy Observatory, which is operated by the Association of
Universities for Research in Astronomy, Inc. under cooperative
agreement with the National Science Foundation.}

\altaffiltext{2}{Jet Propulsion Laboratory, California Institute of
Technology, Mail Stop 169-327, Pasadena, CA 91109 USA [{\tt email:
stern@zwolfkinder.jpl.nasa.gov}]}

\altaffiltext{3}{Department of Physics and Astronomy, The Johns Hopkins
University, Baltimore, MD 21218 USA}

\altaffiltext{4}{Osservatorio Astronomico di Trieste, via G.B. Tiepolo
11, I-34131, Trieste, Italy}

\altaffiltext{5}{Institute of Geophysics and Planetary Physics,
Lawrence Livermore National Laboratory, L-413, Livermore, CA 94550 USA}

\altaffiltext{6}{Physics Department, University of California at Davis,
Davis, CA 95616 USA}

\altaffiltext{7}{European Southern Observatory, Karl-Schwarzschildstr.
2, D-85748, Garching, Germany}

\altaffiltext{8}{Department of Astronomy, University of Florida,
P.O. Box 112055, Gainesville, FL 32611 USA}

\altaffiltext{9}{Department of Physics and Astronomy, University of
Pittsburgh, Pittsburgh, PA 15260 USA}

\altaffiltext{10}{Department of Astronomy, University of California at
Berkeley, Berkeley, CA 94720 USA}

\altaffiltext{11}{National Optical Astronomy Observatory, 950 North
Cherry Avenue, Tucson, AZ 85719 USA}


\altaffiltext{12}{W.~M.~Keck Observatory, 65-1120 Mamalahoa Highway,
Kamuela, HI 96743 USA}

\begin{abstract}

We present our first results on field X-ray sources detected in a deep,
184.7~ks observation with the Advanced CCD Imaging Spectrometer
(ACIS-I) on the {\it Chandra X-Ray Observatory}.  The observations
target the Lynx field ($\alpha_{\rm J2000} = 08^h48^m, \delta_{\rm
J2000} =$ +44\deg54\arcmin) of SPICES, the Spectroscopic Photometric
Infrared-Chosen Extragalactic Survey, which contains three known
X-ray-emitting clusters at redshifts of $z=0.57, 1.26$, and 1.27.  Not
including the known clusters, in the 17\arcmin$\times$17\arcmin\ ACIS-I
field we detect 132 sources in the 0.5$-$2~keV (soft) X-ray band down
to a limiting flux of $\approx 1.7\times10^{-16} \ergcm2s$ and 111
sources in the 2$-$10~keV (hard) X-ray band down to a limiting flux of
$\approx 1.3\times10^{-15} \ergcm2s$.  The combined catalog contains a
total of 153 sources, of which 42 are detected only in the soft band
and 21 are detected only in the hard band.  Confirming previous {\it
Chandra} results, we find that the fainter sources have harder X-ray
spectra, providing a consistent solution to the long-standing `spectral
paradox'.

From deep optical and near-infrared follow-up data, 77\% of the X-ray
sources have optical counterparts to $I = 24$ and 71\% of the X-ray
sources have near-infrared counterparts to $K_s = 20$.  Four of the 24
sources in the near-IR field are associated with extremely red objects
(EROs; $I - K_s \geq 4$).  We have obtained spectroscopic redshifts with
the Keck telescopes of 18 of the Lynx {\it Chandra} sources.  These
sources comprise a mix of broad-lined active galaxies/quasars,
apparently normal galaxies, and two late-type Galactic dwarfs.
Intriguingly, one Galactic source is identified with an M7 dwarf
exhibiting non-transient, hard X-ray emission.

Thirteen of the {\it Chandra} sources are located within regions for
which we have {\it Hubble Space Telescope} imaging.  Nine of the
sources are detected, showing a range of morphologies:  several show
compact cores embedded within diffuse emission, while others are
spatially extended showing typical galaxy morphologies.  Two of the
{\it Chandra} sources in this subsample appear to be associated with
mergers.

We briefly review non-AGN mechanisms to produce X-ray emission and
discuss properties of the Lynx {\it Chandra} sample in relation to
other samples of X-ray and non-X-ray sources.

\end{abstract}


\section{Introduction}

The launch of the {\it Chandra X-Ray Observatory} on 1999 July 23
opened a new era in our study of the X-ray universe.  With its
revolutionary mirror assembly design, {\it Chandra} provides $\approx
0\farcs5$ resolution imaging with $\approx$ 1\arcsec\ astrometry over
the $\sim 0.1 - 10$~keV range \markcite{Weisskopf:96}(Weisskopf, {O'dell}, \&  {van~Speybroeck} 1996).  Furthermore, the
deepest {\it Chandra} observations to date are $\sim 50$ times more
sensitive than the deepest pre-{\it Chandra} observations in the $0.5 -
2$~keV soft X-ray band \markcite{Hasinger:98}(\eg Hasinger {et~al.} 1998) and more than two
orders of magnitude more sensitive than the deepest pre-{\it Chandra}
observations in the $2 - 10$~keV hard X-ray band \markcite{Ueda:99,
Fiore:99}(\eg Ueda {et~al.} 1999; Fiore {et~al.} 1999).  At this resolution and depth, {\it Chandra} has resolved $>
90$~\%\ of the soft X-ray background and $\approx$~80\%\ of the hard
X-ray background into discrete sources \markcite{Mushotzky:00,
Barger:01, Giacconi:01, Hornschemeier:01, Tozzi:01}(\eg Mushotzky {et~al.} 2000; Barger {et~al.} 2001; Giacconi {et~al.} 2001; Hornschemeier {et~al.} 2001; Tozzi {et~al.} 2001).  {\it Chandra}'s
imaging resolution is also well-matched to the typical deep
optical/near-infrared imaging resolutions, providing for the first time
largely unambiguous identifications of the X-ray hosts.  This affords
an exciting new opportunity to study X-ray host identifications,
thereby aiding our understanding of what powers the X-ray background.
In particular, even heavily obscured active galactic nuclei (AGN) are
expected to be luminous and largely unattenuated in the hard X-ray
band:  the great hope is that studying the resolved hard X-ray
background can provide an unbiased picture of the history of
accretion-driven energy production in the Universe.

With the goal of simultaneously studying both the X-ray background and
the properties of intracluster media at high redshift, we obtained a
deep, 184.7~ks exposure with the Advanced CCD Imaging Spectrometer
(ACIS-I) on board {\it Chandra} of three spatially-proximate
high-redshift galaxy clusters in the Lynx field of the SPICES survey
\markcite{Eisenhardt:01}(Eisenhardt {et~al.} 2001, see \S4 below).  All
three clusters were identified in the {\it ROSAT} Deep Cluster Survey
\markcite{Rosati:95, Rosati:98}(RDCS; Rosati {et~al.} 1995, 1998) and
reside within a single ACIS-I field.  Two are at $z > 1$, among the
most distant X-ray-emitting galaxy clusters known:  RX~0848+4453 at $z
= 1.27$ \markcite{Stanford:97}(Stanford {et~al.} 1997) and RX~0849+4452
at $z = 1.26$ \markcite{Rosati:99}(Rosati {et~al.} 1999).  The third
cluster is at $z=0.57$ \markcite{Rosati:98}(RX~0848+4456; Rosati
{et~al.} 1998).  At 184.7~ks, our {\em Chandra} observation is among
the deepest observations of the X-ray sky taken to date.  Analysis of
the diffuse X-ray emission from the intracluster media of the two $z >
1$ clusters is presented in \markcite{Stanford:01}Stanford {et~al.}
(2001) and suggests little evolution in the X-ray luminosity $-$
temperature ($L_X - T_X$) relation to $z \approx 1.3$.  Analysis of the
diffuse X-ray emission from the $z = 0.570$ cluster is presented in
\markcite{Holden:01}Holden {et~al.} (2001) and suggests that 10$-$20\%
of the X-ray emission actually comes from a group at $z=0.543$ along
the line of sight, showing the importance of detailed and
multiwavelength studies of galaxy clusters when using them as probes of
cosmological mass functions.  Observations of the 153 X-ray point
sources are the subject of the remainder of this paper.

To aid in follow-up observations, this manuscript details and provides
a catalog of the unresolved X-ray sources we identify from the deep
17\arcmin\ $\times$ 17\arcmin\ ACIS-I image (\S~2).  In \S~3 we discuss
results from the X-ray data alone, including the hardness ratio and the
fraction of the X-ray background which is resolved from these
observations.  SPICES optical ($BRIz$) images cover the central
16\arcmin $\times$ 16\arcmin\ of this field and SPICES near-infrared
($JK_s$) images cover the central 5\farcm6$\times$ 5\farcm6 of this
field:  we present the results of our cross-correlation of the
optical/near-infrared imaging data with the X-ray catalog in \S~4.  The
18 redshifts we have obtained for Lynx {\em Chandra} sources are also
discussed in \S~4.  Section~5 presents our first spectroscopic results
and \S~6 highlights some initial results from this work.  We summarize
our conclusions in \S~7.   An additional, interesting X-ray source in
this field, CXO52, is the subject of \markcite{Stern:02}Stern \etal
(2002). Unless otherwise stated, all magnitudes presented are in the
Vega system.  Cosmology-dependent parameters are calculated for two
models:  an Einstein-de~Sitter universe consistent with previous work
in this field  ($H_0 = 50~h_{50}~ \kmsMpc$, $\Omega_M = 1$, and
$\Omega_\Lambda = 0$) and the dark energy universe favored by recent
work on high-redshift supernovae and fluctuations in the cosmic
microwave background \markcite{Riess:01}($H_0 = 65~\kmsMpc$, $\Omega_M
= 0.35$, and $\Omega_\Lambda = 0.65$; \eg Riess {et~al.} 2001).

\section{X-Ray Data:  Observations and Source Detection}

\subsection{Observations and data reduction }

ACIS-I observations of the Lynx field were obtained on 2000 May 3
(65~ks; OBS-ID 1708) and 2000 May 4 (125~ks; OBS-ID 927).  The aimpoint
for the observations was $\alpha_{\rm J2000} = 08^h48^m54.79^s$,
$\delta_{\rm J2000} =$ +44\deg54\arcmin32\farcs9 and both exposures
were obtained in the faint mode when ACIS was at a temperature of
$-120\deg$~C.  The Galactic absorbing column for this field is $N_H = 2
\times 10^{20}~ {\rm cm}^{-2}$.  The position angle of the observations
was 258.45\deg.

The data were reduced and analyzed using the {\it Chandra} Interactive
Analysis of Observations (CIAO) software (release V1.1, see {\tt
http://asc.harvard.edu/ciao}).  The Level~1 data were processed using
the quantum efficiency uniformity file for $-120\deg$~C, which corrects
the effective area for loss due to charge transfer inefficiency.
The correction compensates for the loss of events far from the readout,
especially at high energies, and is relevant when fitting the total
spectrum in the largest energy range.  The data were filtered to include
only the standard event grades 0, 2, 3, 4, and 6.  All bad columns were
removed by eye, as were the columns close to the border of each node,
since the grade filtering is not efficient in these columns.  We also
identified flickering pixels in the images in the chip coordinates:  for
each chip, we checked photon arrival times for all pixels on which more
than two photons were detected.  If two or more photons were detected
within an interval shorter than 7 seconds, they were considered flickering
pixels and excluded from further analysis.  As the brightest source
(CXO39) has a count rate of only 1 photon every $\approx 100$~seconds,
this procedure should not affect our source catalog.  The removal of
columns and pixels reduces slightly the effective area of the detector,
the result of which has been included when calculating the soft and hard
exposure maps.

Time intervals with background rates larger than $3 \sigma$ over the
quiescent value ($\simeq 0.30$ counts s$^{-1}$ per chip in the $0.3 -
10$ keV band) were removed.  This procedure gave 60.7~ks of effective
exposure out of the first observation, and 124~ks out of the second,
for a total of 184.7~ks.  The two observations are almost coincident on
the sky, so that the total coverage is 298 arcmin$^2$.  The two event
files were merged into a single file, which was used in the reduction
and analysis process.  The astrometry of the two observations,
contiguous in time, was perfectly consistent, as verified by
registering the positions of the brightest sources on top of each
other.  We did not include data from the chips S-2 and S-3 in the
following analysis, since the point spread function (PSF) is very
broadened ($\simeq 15$ arcsec) and they would add a very little
effective area even for high fluxes.  Fig.~\ref{Xcolorfig} shows the
final, combined X-ray data, with a false-color mapping derived from the
{\it Chandra} energy bands.

\subsection{Source detection}

A soft band from 0.5$-$2~keV and a hard band from 2$-$7~keV were used
in the detection process to select a soft and hard source lists, as
detailed below.  The hard band detection window was cut at 7~keV since
above this energy the effective area of {\it Chandra} is decreasing,
while the instrumental background is rising, giving very inefficient
detection of sky photons.  X-ray count rates in the above bands were
used to calculate X-ray fluxes in a 0.5$-$2~keV soft band and
2$-$10~keV hard band, assuming an average X-ray spectrum.

We ran Source Extractor \markcite{Bertin:96}(Bertin \& Arnouts 1996) on
the summed 0.5$-$7 keV image binned by a factor of two (one binned
pixel corresponds to 0.982\arcsec $\times$ 0.982\arcsec).  Since we are
far from being background limited, the detection efficiency proved to
be much higher for the 0.5$-$7~keV image rather than the soft and hard
bands separately.  \markcite{Giacconi:01}Giacconi \etal (2001) use the
same procedure for their analysis of the {\it Chandra} Deep Field-South
(CDF-S).  Source Extractor parameters were chosen after extensive
simulations, and we adopted an internal detection threshold
signal-to-noise ($S/N$) ratio of 2.4, with a Gaussian filter FWHM of
1\farcs5 and a minimum area of 5 pixels.  Since Source Extractor is not
tailored to work with a very low background, it gives a wrong estimate
of the internal background.  To tackle this problem, we built a
smoothed map of the background computed from the data themselves in the
corresponding band, after the removal of the sources down to a very low
threshold.  This background map was used as a weight (variance) map to
define the detection threshold.  This modified detection algorithm is
several orders of magnitude faster than the wavelet detection algorithm
of \markcite{Rosati:95}Rosati \etal (1995) or the {\tt WAVDETECT}
algorithm in the CIAO software \markcite{Freeman:02}(Freeman \etal 2002).

Table~1 presents the catalog of $0.5 - 7$~keV point sources identified
in the combined image, selected as follows.  We first measured the
signal-to-noise ($S/N$) ratio of all the candidate detections in
circular extraction apertures.  A radius $R_S = 2.4 \times {\rm
FWHM}_{\rm PSF}$ was used, where ${\rm FWHM}_{\rm PSF}$ is the modeled
PSF full-width at half-maximum, determined as a function of off-axis
angle to reproduce the broadening of the PSF.  A minimum $R_S =
5\arcsec$ was used in the central regions.  Sources with $S/N > 2.1$
within the extraction area were considered detections.  This limit was
chosen to proved a census of X-ray sources to faint flux levels, where
the choice of Source Extractor parameters has already significantly
reduced the number of spurious detections.  We stress as a compromise
between providing a census of X-ray sources to faint levels and
limiting the contamination from false detections.  We stress that the
condition of having $S/N > 2.1$ in the extraction area corresponds to a
high significance detection -- the faintest detected sources have more
than 10 counts.  These catalogs, obtained from the Source Extractor run
and the a-posteriori $S/N$ cut, typically include less than five
spurious sources as tested with simulations with the Model of {\em
AXAF}\footnotemark\ Response to X-rays (MARX) software
\markcite{Tozzi:01}(for a detailed description of the simulations, see
Tozzi {et~al.} 2001).  After performing this procedure on the soft and
hard images separately, we obtained the soft and hard catalogs.  A
combined catalog was then produced by matching the two.  Source counts
were measured with simple aperture photometry within $R_S$.  The
background for each source was calculated locally within an annulus of
inner radius $R_S + 2\arcsec$ and an outer radius of $R_S +
12\arcsec$.  Numerical simulations show that our aperture photometry
systematically underestimates the net counts by 4\%.  For the following
analysis, we have applied this aperture correction.

\footnotetext{The {\em Chandra X-ray Observatory} was formerly known as
{\em AXAF}, the {\em Advanced X-ray Astrophysics Facility}.}

Finally, we transformed the net count rate into energy flux in the soft
0.5$-$2~keV band and in the hard 2$-$10~keV band.  The conversion
factors used were $(4.52 \pm 0.3) \times 10^{-12}~ {\rm erg}~ {\rm
cm}^{-2}~ {\rm count}^{-1}$ in the soft band and $(2.79 \pm 0.3) \times
10^{-11}~ {\rm erg}~ {\rm cm}^{-2}~ {\rm count}^{-1}$ in the hard band,
assuming an absorbing column of $2 \times 10^{20}$ cm$^{-2}$ (Galactic
value) and a photon index\footnotemark\ $\Gamma = 1.4$.  The
uncertainties in the conversion factors reflect the range of possible
values for the photon index, $\Gamma = 1.1 - 1.8$.  As suggested by the
spectral analysis of the stacked spectra, these values are
representative of our sample (see below).  Since the conversion factors
were computed at the aimpoint, where the effective area of {\it
Chandra}/ACIS is at its maximum, the count rate of each source was
corrected for vignetting.  Namely, the net count rate was multiplied by
the ratio of the exposure map at the aimpoint to the value of the
exposure map averaged within the extraction region.  Such a correction
was done in the soft and hard bands separately.

\footnotetext{The number of X-ray photons $N(E)$ detected as a function
of energy $E$ is assumed to follow a power law distribution, $N(E) dE
\propto E^{-\Gamma} dE$, where $\Gamma$ is the photon index.}

We detect 132 sources in the 0.5$-$2~keV (soft) X-ray band down to a
2.1$\sigma$ limiting flux of $S_{0.5-2} \approx 1.7\times10^{-16}
\ergcm2s$ and 111 sources in the 2$-$10~keV (hard) X-ray band down to a
2.1$\sigma$ limiting flux of $S_{2-10} \approx 1.3\times10^{-15}
\ergcm2s$.  Only the point sources are considered here.  As mentioned
above, the diffuse galaxy cluster emissions are explored by
\markcite{Stanford:01}Stanford {et~al.} (2001) and
\markcite{Holden:01}Holden {et~al.} (2001).  Of the total catalog of
153 sources, 42 (27\%) are detected only in the soft band and 21 (14\%)
are detected only in the hard band.  These results are very similar to
the fractions derived by \markcite{Tozzi:01}Tozzi {et~al.} (2001) for
their 300~ks ACIS map of the CDF-S:  referred to the total sample, they
find 26\%\ of the sources are detected only in the soft band and
11\%\ of the sources are detected only in the hard band.

Astrometric positions were initially determined from the ACIS-I aspect
solution.  Following comparison of X-ray source positions to the deep,
ground-based imaging discussed in \S4, we refined the absolute pointing
of the X-ray data with an offset of 1\farcs09 west, 1\farcs35 south,
and no rotation.  These offsets were determined by matching X-ray
sources with greater than 25 total counts ($0.5 - 7$~keV) to $17.5 < I
< 22.5$ sources.  The positions presented in Table~2
therefore match the astrometric frame of the SPICES survey.

\section{X-Ray Data:  Results}

\subsection{Resolving the X-ray background}

To compute the number counts in the soft and hard bands, we first
calculate the sky coverage as a function of the flux.  The effective
sky coverage of a given flux is defined as the area on the sky where a
soft (or hard) source can be detected with a $S/N > 2.1$ in the
extraction region defined above.  We have performed extensive
simulations with MARX to show this procedure is accurate within a few
percent \markcite{Tozzi:01}(\eg see Tozzi {et~al.} 2001).  The same
procedure has been used for the CDF-S \markcite{Giacconi:01}(Giacconi
{et~al.} 2001).

In Fig.~\ref{fig_NSsoftXRB} we present the soft X-ray source counts
derived from our observations.  For comparison, we also show source
counts from the deep {\it ROSAT} observations of the Lockman Hole
\markcite{Hasinger:98}(Hasinger {et~al.} 1998), extrapolation of the
source counts to fainter fluxes from fluctuation analysis of the {\it
ROSAT} data \markcite{Hasinger:93}(Hasinger {et~al.} 1993), and recent,
deep {\it Chandra} soft X-ray source counts from
\markcite{Mushotzky:00}Mushotzky {et~al.} (2000).  We find the {\it
Chandra} results in excellent agreement with {\it ROSAT} in the region
of overlap $S_{0.5-2} >10^{-15}$ \ergcm2s.  The {\it Chandra} data
extend the results to $2 \times 10^{-16}$ \ergcm2s.  A
maximum-likelihood power-law fit to the soft X-ray $\log N - \log S$
data finds $$N (> S_{0.5-2}) = 320 \pm 80 \left( { S_{0.5-2} \over 2
\times 10^{-15} \ergcm2s} \right) ^{-0.82 \pm 0.10}.$$ Confidence
contours for the slope and the normalization of the soft counts are
shown in the upper right corner of Fig.~\ref{fig_NSsoftXRB}, compared
with the results of \markcite{Mushotzky:00}Mushotzky {et~al.} (2000).
Note that here we used an average slope to be consistent with the
average spectrum of our sample of sources.

In Fig.~\ref{fig_NShardXRB} we present the hard X-ray source counts
derived from our observations down to a flux limit of $2 \times
10^{-15}$ \ergcm2s.  At bright levels source counts from the {\it
Advanced Satellite for Cosmology and Astrophysics} ({\it ASCA}) and
{\it BeppoSAX} are shown \markcite{Giommi:98, dellaCeca:99, Ueda:99}(Giommi {et~al.} 1998; {della~Ceca} {et~al.} 1999; Ueda {et~al.} 1999).  At
faint levels we show recent, deep {\it Chandra} hard X-ray source
counts from \markcite{Mushotzky:00}Mushotzky {et~al.} (2000).  A maximum-likelihood power-law fit
to the hard X-ray $\log N - \log S$ data finds $$N (> S_{2-10}) = 1400
\pm 200 \left( { S_{2-10} \over 2 \times 10^{-15} \ergcm2s} \right)
^{-1.12 \pm 0.17}.$$ Confidence contours are again provided, as per
Fig.~\ref{fig_NSsoftXRB}.  Our hard-band $\log N - \log S$ results are
consistent with deep observations of the Lockman Hole obtained with the
{\it X-Ray Multi-Mirror-Newton} satellite \markcite{Hasinger:01}({\it
XMM-Newton};  Hasinger {et~al.} 2001).  Our normalization of the hard-band $\log
N - \log S$ relation is 1$\sigma$ higher than the results for the CDF-S
who find a normalization to the above relation of $K_{15} = 1150 \pm
150$ where $N(>S_{2-10}) = K_{15} (S_{2-10} / 2 \times 10^{-15}
\ergcm2s)^{-\alpha}$ \markcite{Tozzi:01}(Tozzi {et~al.} 2001).   In both the soft and the
hard X-ray bands, the Lynx-field source counts are slightly lower at
each X-ray flux than those determined for SSA13 \markcite{Mushotzky:00}(Mushotzky {et~al.} 2000).
The difference likely reflects cosmic variance due to large scale
structure on size scales similar to the {\it Chandra} field-of-view.
In particular, the SSA13 results analyze only one ACIS chip, thus
sampling one-quarter the area of the results discussed here.  We note
that four of the deepest 16\arcmin\ $\times$ 16\arcmin\ {\it Chandra}
fields (CDF-S, HDF-N, the Lockman Hole, and Lynx) find consistent
normalizations to the $\log N - \log S$ relation, while SSA13 and the
MS1137 fields give 40\%\ higher normalizations (Tozzi, in preparation)

What fraction of the hard X-ray background is resolved by the deep Lynx
{\it Chandra} map?  Fig.~\ref{fig_hardXRB2} presents the integrated
contribution of resolved hard X-ray sources as a function of X-ray
flux.  In the range $(1 - 100) \times 10^{-15} \ergcm2s$, the
integrated hard X-ray flux density in the 2$-$10~keV band is $(1.2 -
1.3 \pm 0.1) \times 10^{-11} \cgsdeg2$, depending on the assumed
average power law ($\Gamma = 1.6 - 1.4$, respectively).  Including the
hard-band $\log N - \log S$ relation derived by \markcite{dellaCeca:99}{della~Ceca} {et~al.} (1999)
for a wider-area, brighter {\it ASCA} survey, we find that $(1.5 \pm
0.2) \times 10^{-11} \cgsdeg2$ of the hard X-ray background is resolved
to a flux limit of $10^{-15} \ergcm2s$.  As seen in
Fig.~\ref{fig_hardXRB2}, this is close to the {\it High Energy
Astronomy Observatory-1} ({\it HEAO-1}) value of $1.6 \times 10^{-11}
\cgsdeg2$ \markcite{Marshall:80}(Marshall {et~al.} 1980), but $20-40$\%\ lower than the more
recent determinations from {\it BeppoSAX} and {\it ASCA}
\markcite{Vecchi:99, Ishisaki:99}(\eg Vecchi {et~al.} 1999; Ishisaki {et~al.} 1999).  For comparison, from the first
300~ks observation of the CDF-S, \markcite{Tozzi:01}Tozzi {et~al.} (2001) resolve $(1.45 \pm
0.15) \times 10^{-11} \cgsdeg2$ of the hard X-ray background while
\markcite{Mushotzky:00}Mushotzky {et~al.} (2000) resolve $\approx 1.8 \times 10^{-11} \cgsdeg2$
in their 101~ks observation of SSA13.  We conclude that, given the
uncertainty on the value of the total background, a fraction between
$20$\% and $40$\% of the hard X-ray background is still unresolved in
our data set.

\subsection{Hardness ratio}

A long-standing problem in X-ray astrophysics has been that the
spectral slope of the X-ray background is significantly steeper than
the spectral slopes of the first (brightest) soft X-ray sources which
were identified.  The majority of these bright sources are identified
with unobscured AGN \markcite{Schmidt:98}(\eg Schmidt {et~al.} 1998).
This problem, the so-called ``spectral paradox,'' implies that a new
population of fainter sources with harder X-ray indices exists:
obscured AGN are the likely culprit.  {\it Chandra} has now identified
this population.

Fig.~\ref{fig_softhard} illustrates the hardness ratio $HR \equiv
(H-S)/(H+S)$ for the Lynx sample, where $H$ and $S$ are the net counts
in the hard ($2-7$~keV) and soft ($0.5-2$~keV) bands, respectively.
The hardness ratio is plotted against the hard X-ray flux.  There are
21 sources, corresponding to 14\% of the total combined sample, which
are detected only in the hard band (plotted at $HR=1$) and 42 which are
detected only in the soft band (not plotted).  A population of hard
sources, largely absent at bright fluxes, clearly emerges at faint
fluxes.  These results are analogous with the ones found in the CDF-S
\markcite{Giacconi:01}(Giacconi {et~al.} 2001).  We note that in the
Lynx field there are three bright outliers in the hardness ratio--flux
plane, in the upper right corner.  These are very absorbed sources
($10^{22} < N_H < 10^{23}$ cm$^2$).  In particular, the brightest one,
CXO12, dominates the spectrum of the bright end of our sample, giving
an effective $\Gamma \simeq 1.3$ already for fluxes $S_{2-10}\simeq
10^{-13}$ erg cm$^{-2}$ s$^{-1}$.  CXO12 is identified with the $z =
0.9$ Type~2 quasar AX~J08494+4454 reported by \markcite{Ohta:96}Ohta
\etal (1996) and \markcite{Nakanishi:00}Nakanishi \etal (2000) from
{\it ASCA} observations of the Lynx field.  Of the other two hard,
bright X-ray sources which stand out in Fig.~\ref{fig_softhard}, only
one (CXO36) is within the area imaged optically in the current work.
Optical spectra from these other two sources are not discussed in the
current work.

Fig.~\ref{fig_allsources} presents the average spectral index of the
population, determined from the stacked spectrum of all 154 sources in
the combined sample.  As a background, we used the stacked spectrum of
all the background regions extracted around each source.  The resulting
photon files were scaled by the ratio of the total area of extraction
of the sources and the corresponding area for the background.  This
procedure guarantees a correct background subtraction despite the
non-uniformities of the instrumental background across ACIS-I.  The
response matrices for the stacked spectra were obtained from the
counts-weighted average of the matrices for individual sources.  XSPEC,
the X-ray spectral fitting package \markcite{Arnaud:96}(Arnaud 1996), was used to
compute the slope of a power law spectrum with local $N_H$ absorption
in the energy range $0.5-7$~keV.  We excluded bins below $0.5$~keV
because the calibration is still uncertain below this energy.  We
obtained a photon index $\Gamma$ of $1.36 \pm 0.04 $ and a column
density $N_H = (2.3 \pm 1.4) \times 10^{20}$ cm$^{-2}$ which is
consistent with the Galactic value (errors refer to the 90\% confidence
level).  The results of the spectral fits are shown in
Fig.~\ref{fig_allsources}.  We find an average spectrum for the
detected sources is consistent with the average shape of the hard
background ($\langle \Gamma \rangle \simeq 1.4$).


\section{Optical/Near-Infrared Imaging}

\subsection{Summary of survey observations }

The Lynx field is one of four fields which comprise SPICES
\markcite{Eisenhardt:01}(Eisenhardt {et~al.} 2001).  This long-term project was conceived as the next
generation analog to the $I$-selected Canada-France Redshift Survey
\markcite{Lilly:96}(CFRS; Lilly {et~al.} 1996):  by studying a sample of $K$-selected
faint field galaxies over a larger, 100 arcmin$^2$ field, we are
attempting to address the history of galaxy formation and evolution
out to $z \simeq 1.5 - 2$ in a manner which is less-biased by large
scale structure and recent star formation.  In particular, selecting
galaxies at $2 \mu$m should essentially translate to selecting by mass,
thereby providing a powerful discriminant of theories of galaxy assembly
\markcite{Kauffmann:98}(\eg Kauffmann \& Charlot 1998).  Four unremarkable but well-studied fields
were chosen spread across the sky for efficient follow-up throughout the
year.  Details of the motivation and imaging observations are presented in
\markcite{Eisenhardt:01}Eisenhardt {et~al.} (2001).  We briefly summarize the Lynx observations here.

Optical $BRIz$ imaging was obtained using the Kitt Peak National
Observatory 4~m Mayall telescope with its Prime Focus CCD imager
(PFCCD) equipped with a thinned AR coated $2048 \times 2048$ Tektronics
CCD.  This configuration gives a 16\arcmin $\times$ 16\arcmin\ field of
view with 0\farcs47 pix$^{-1}$.  The filters used were a Harris
$B$-band ($\lambda_c = 4313$ \AA; $\Delta \lambda = 1069$ \AA), Harris
$R$-band ($\lambda_c = 6458$ \AA; $\Delta \lambda = 1472$ \AA), Harris
$I$-band ($\lambda_c = 8204$ \AA; $\Delta \lambda = 1821$ \AA), and an
RG850 long-pass $z$-band filter.  For the $I$ and $z$ imaging, the CCD
was operated using ``short scan,'' where the CCD was mechanically
displaced while its charge is shifted in the opposite direction to
reduce fringing to very low levels.  The combined, processed images
reach limiting Vega magnitudes of 26.9 ($B$), 25.4 ($R$), 24.6 ($I$),
and 24.2 ($z$), where these numbers represent 3$\sigma$ limits in
3\arcsec\ diameter apertures.  The corresponding AB magnitude limits
are 26.8, 25.6, 25.1, and 24.7, respectively.  The seeing for the
summed images ranges from 1\farcs2 ($B$) to 1\farcs0 ($z$).  Data
reduction followed standard techniques \markcite{Eisenhardt:01}(for details,
see Eisenhardt {et~al.} 2001).

Near-infrared $JK_s$ imaging was obtained at the KPNO 4~m with its
Infrared Imager \markcite{Fowler:88}(IRIM; Fowler {et~al.} 1988) equipped with a Rockwell
International NICMOS~3 256 $\times$ 256 HeCdTe array giving 0\farcs6
pix$^{-1}$.  Four pointings were obtained to cover a $5\farcm6 \times
5\farcm6$ area.  The average exposure time in the fully sampled regions
of the resulting mosaic was 10.14~ks in $J$ ($\lambda_c = 1.14 \mu{\rm
m}; \Delta \lambda = 0.29 \mu{\rm m}$) and 8.49~ks in $K_s$ ($\lambda_c
= 2.16 \mu{\rm m}; \Delta \lambda = 0.33 \mu{\rm m}$).  The seeing in
the mosaics is 1\farcs2 $-$ 1\farcs3.  Data reduction followed standard
techniques \markcite{Eisenhardt:01}(for details, see Eisenhardt {et~al.} 2001).  Calibrations of
the optical and near-infrared images onto the Landolt and CIT systems
were obtained using observations of \markcite{Landolt:92}Landolt (1992) and UKIRT
standard stars \markcite{Hawarden:01}(Hawarden {et~al.} 2001), respectively. The $J$-band images
reach a 3$\sigma$ limiting magnitude of 22.9 (Vega) in a
3\arcsec\ diameter aperture.  The corresponding depth for the $K_s$
images is $K_s = 21.4$.

We used Source Extractor \markcite{Bertin:96}(release V2.1.6; Bertin \& Arnouts 1996) to create
independent catalogs of sources selected from the $B$-, $I$-, and
$K_s$-band images.  Photometry was generated for 3\farcs0 diameter
apertures and evaluated in the same apertures for all six coaligned
bands.

\subsection{Optical/near-infrared counterparts}

Fig.~\ref{fig_offsets} shows the X-ray-to-optical positional offsets
after the X-ray positions have been shifted by 1\farcs09 west and
1\farcs35 south (see \S 2.2).  Based on these results, we matched the
optical/near-infrared and X-ray source catalogs using a 1\farcs5 radius
aperture for host identification.  Table~2 presents the
optical/near-infrared properties of the {\it Chandra} sources listed in
Table~1.  As indicated in the table, for some sources, primarily
fainter X-ray sources for which the X-ray position is more uncertain,
we have used a larger ($1\farcs5 - 2\farcs3$) match radius.
Non-detections are listed with an ellipsis (``...'') while sources with
bad photometry, due either to source saturation or contamination from
the stellar bleed trail of a nearby, bright source, are listed as
99.00.  Only those sources within the smaller near-infrared field are
tabulated in the near-infrared photometry columns.  Sources from
Table~1 not within the optical field are not included in Table~2.  Only
104 of the X-ray sources are within the optical image field of view.
Of these, 80, or 77\%, have $I < 24$ (Vega) identifications.  The
infrared field of view covers 24 of the X-ray sources, of which 17, or
71\%, have $K_s < 20$ (Vega) identifications.  When we restrict
ourselves to {\it Chandra} sources with signal-to-noise ratios greater
than 3, the absolute numbers of X-ray sources in each region drops by
$\sim 30$\% but the fraction of sources with $I$ and/or $K_s$
identifications does not change substantially.

From the SPICES imaging we find that the surface density of sources to
$K_s = 20$ is $\approx$ 15 arcmin$^{-2}$, so the probability of a
chance coincidence of an X-ray source within 1\farcs5 of a $K_s \leq
20$ source is $\approx 3$\%.  We therefore expect none of the 17 $K <
20$ identifications to be spurious.  The surface density of $I < 24$
(Vega) sources is $\approx$ 23.5 arcmin$^{-2}$ so the probability of a
chance juxtaposition of an X-ray source with an $I \leq 24$ source with
our 1\farcs5 match radius is $\approx 5$\%.  We therefore expect that
up to five of the faint host identifications in Table~2 to be spurious.
Note that most of the optical identifications are in fact much brighter
than $I = 24$, and are thus less likely to be spurious.

In Fig.~\ref{fig_IRcolor} we present color-magnitude diagrams for
sources detected in our near-infrared imaging field.  X-ray sources are
indicated with larger symbols.  We see that for a given optical
magnitude, the X-ray source hosts tend to be red in optical-to-near-IR
color, hugging the so-called {\em red envelope}.  Several of these
sources meet the standard identity criteria of extremely red objects
\markcite{Graham:96}(EROs; \eg Graham \& Dey 1996):  one source has $R - K > 6$ and four
sources have $I - K > 4$.  We know from previous X-ray surveys that
many bright, early-type galaxies at moderate redshift are X-ray
luminous; {\it Chandra}-identified EROs may represent the
higher-redshift tail of that population.  Indeed, \markcite{Cowie:01}Cowie {et~al.} (2001)
suggest that identifying optically-faint, hard X-ray sources may prove
an efficient strategy to locate luminous, evolved galaxies at high
redshift ($z \simgt 1.4$).  We note that optically-faint, {\em soft}
X-ray sources are also occasionally identified with high-redshift,
early-type galaxies (\eg CXO138; \S~6.4.2).  Highly-obscured sources
are also expected to have red colors and may represent some fraction of
extremely red X-ray sources.  Contrary to that expectation, the $z =
3.70$ Type~II quasar in the CDF-S identified by \markcite{Norman:01}Norman {et~al.} (2001) is
relatively blue in optical-to-near-IR color: $R-K = 2.56$ and $I-K =
1.68$.  A smaller subset of the X-ray sources have identifications
which have blue optical-to-near-IR colors.  These sources tend to be
associated with broad-lined, bright quasars.

\subsection{{\it Hubble Space Telescope} morphologies}

Thirteen of the {\it Chandra} sources in the Lynx field are located in
regions for which we have images from the Wide Field Planetary Camera~2
\markcite{Trauger:94}(WFPC2; Trauger {et~al.} 1994) on the {\it Hubble Space Telescope} ({\it
HST}; Fig.~\ref{fig_hst}).  These observations, sampling several
independent observing programs, primarily target galaxy clusters in the
Lynx field and range from 2 to 10 orbits (4.8 to 27~ks) through the
F702W (12~ks; RX~J0848+4456 region) and F814W (all other imaging)
filters.  The observations are described in \markcite{vanDokkum:01}van Dokkum {et~al.} (2001),
\markcite{Holden:01}Holden {et~al.} (2001), and Wu \etal (in preparation).

Only nine ($69$\%) of the sources are detected in the {\it HST} images,
lower than expected for a larger sample given that $77$\% of the full
catalog have $I < 24$ identifications.  This fraction is also lower
than the 92\%\ of CDF-S X-ray sources with optical counterparts in the
deeper {\it HST}/WFPC2 study of \markcite{Schreier:01}Schreier {et~al.} (2001), which reaches a
detection limit of $I = 28.2$.  Of the four Lynx non-detections (CXO46,
CXO127, CXO135, and CXO136), three are detected in the X-ray with
signal-to-noise ratios $S/N \leq 3$.  The nine detections show a range
of morphology.  CXO42 and CXO49 are unresolved spatially, though both
show faint, surrounding diffuse emission.  CXO41 also has a very strong
core, embedded in extended emission suggestive of a tidal tail or
spiral arm.  CXO44 and CXO124 also appear to be associated with merging
systems, while the remaining sources (CXO52, CXO60, CXO122, and CXO205)
are spatially extended.  The faint, soft X-ray source CXO205 appears to
be associated with an early-type galaxy.

\subsection{X-ray-to-optical flux ratio}

Fig.~\ref{fig_Xopt} illustrates the X-ray-to-optical flux ratio of soft
and hard X-ray sources.  Following \markcite{Hornschemeier:01}Hornschemeier {et~al.} (2001), we plot
lines of constant X-ray-to-optical flux ratio using the relation
$$\log\left({{f_x}\over{f_R}}\right) = \log f_x + {R\over{2.5}} +
5.50,$$ derived from the Kron-Cousins $R$-band filter transmission
function.  At bright X-ray fluxes, we plot the results of shallow,
wide-area surveys:  for soft X-rays, we show AGN from the {\it ROSAT}
survey reported by \markcite{Schmidt:98}Schmidt {et~al.} (1998) and for hard X-rays, we show
sources from the {\it ASCA} survey reported by \markcite{Akiyama:00}Akiyama {et~al.} (2000).  At
faint X-ray fluxes we show the results of the {\it Chandra} Lynx
survey reported here and results from the 221.9~ks {\it Chandra} study
of the {\it Hubble} Deep Field North (HDF-N) reported in
\markcite{Hornschemeier:01}Hornschemeier {et~al.} (2001).  The majority of sources have $-1 \simlt
\log(f_x/f_R) \simlt 1$, which, as we see from the shallow surveys, is
typical of AGN.  However, at the fainter fluxes probed by {\it
Chandra}, new populations become apparent.

At faint, soft X-ray fluxes, several sources are seen which are X-ray
underluminous for their optical magnitudes.  Optical spectroscopy
\markcite{Hornschemeier:01, Tozzi:01}(see \S5 and Hornschemeier {et~al.} 2001; Tozzi {et~al.} 2001) reveal many of these
sources to be apparently normal galaxies.  We discuss these sources
further in \S6.4.  Several Galactic M dwarfs are also identified with
$\log(f_x/f_R) \approx -2$, discussed further in \S6.3.  We note that
one late-type Galactic dwarf has been identified in the Lynx field more
than two orders of magnitude brighter in the X-ray bands than the other
Galactic X-ray emitters:  we discuss this interesting source further in
\S~6.3.

At faint, hard X-ray fluxes, a population of sources appears which are
X-ray {\em overluminous} for their optical magnitudes.  They potentially
represent an astronomically interesting new population.  Unfortunately,
most are extremely faint optically and lack spectroscopic
identification.  Extremely high-redshift ($z \simgt 5$) AGN might have
bright X-ray fluxes but faint $R$-band fluxes due to absorption from
the Lyman transitions of hydrogen along our line of sight.  Obscured
AGN, for which high column density absorption near the central black
hole shields both the broad-line region and soft X-ray emission, might
also be overluminous in the hard X-ray relative to their optical
magnitudes.  Several examples of these {\em Type~II quasars} have now
been identified in the deepest {\it Chandra} surveys
\markcite{Norman:01, Dawson:01}(\eg Norman {et~al.} 2001; Dawson {et~al.} 2001).

\section{Spectroscopic Observations}

To date, the SPICES survey has obtained spectroscopic redshifts for 219 of
the 485 $K_s < 20$ sources in the Lynx field.  This work started in February
1997 and has all been done in multislit mode with the Low Resolution
Imaging Spectrometer \markcite{Oke:95}(LRIS; Oke {et~al.} 1995) on the Keck telescopes.
Observations typically used the 150 lines mm$^{-1}$ grating ($\lambda_{\rm
blaze} = 7500$ \AA; $\Delta \lambda_{\rm FWHM} \approx 17$ \AA) and sample
the wavelength range 4000 \AA\ to 1$\mu$m.  Typical slitlet lengths
were $\approx$~20\arcsec\ and we performed $\approx$~3\arcsec\ spatial
offsets between each 1200~s $-1800$~s exposure in order to facilitate
removal of fringing at long wavelength ($\lambda \simgt 7200$ \AA).
Masks were designed to contain sources of comparable $I$-band magnitudes.
For the masks containing the brightest sources, two 1200~s exposures
were sufficient to obtain redshifts for most sources on the mask.
For the masks containing the faintest sources, five 1800~s exposures
were not uncommon and some sources have been on multiple faint source
masks so that they now have $\simgt$~10~hr of spectroscopic integration.

Our infrared selection led to the unintentional spectroscopic targeting
of a handful of X-ray sources prior to our obtaining the {\it Chandra}
map.  On UT 2000 October 01 we observed a mask dedicated to bright {\it
Chandra} sources for 2400~s in photometric conditions.  In total, we
have spectroscopic redshifts for 18 of the Lynx {\it Chandra} sources
currently.  Table~3 summarizes the spectroscopic results
and Fig.~\ref{spectra} presents the spectra.  Fig.~\ref{zhist} shows
the histogram of redshifts for spectroscopically-identified X-ray
sources in this field.

All data reductions were performed using IRAF and followed standard
slit spectroscopy procedures.  We calculated the pixel-to-wavelength
transformation using a NeAr lamp spectrum, generally observed immediately
subsequent to the science observations (RMS variations of 0.6 \AA\ are
typical for the 150 lines mm$^{-1}$ grating), and employed telluric
emission lines to adjust the wavelength zero-point.  The spectra on
photometric nights were flux-calibrated using observations of standard
stars from \markcite{Massey:90}Massey \& Gronwall (1990).

Consistent with previous spectroscopic studies of faint X-ray hosts
\markcite{Fiore:00, Barger:01, Giacconi:01}(\eg Fiore {et~al.} 2000; Barger {et~al.} 2001; Giacconi {et~al.} 2001), the Lynx {\it Chandra}
sources are associated with a wide range of astronomical objects.  We
find two Galactic dwarfs, seven obvious AGN, and nine apparently normal
galaxies.  Of the galaxies, four are early-type and five show evidence
of star formation.

\section{Discussion}

\subsection{Luminosities and Spectral Classification}

Table~3 presents the optical and X-ray luminosities for
the sample of sources for which we have spectroscopic redshifts.
Luminosities have been calculated for two cosmologies:  an
Einstein-de~Sitter (EdS) universe with $H_0 = 50 \kmsMpc$, consistent
with previous work in this field \markcite{Tozzi:01}(\eg Tozzi {et~al.} 2001) and the dark
energy ($\Lambda$) universe favored by recent supernovae and microwave
background observations \markcite{Riess:01}(\eg Riess {et~al.} 2001).

We calculate rest-frame $B$-band luminosities $M_B$ by interpolating
rest-frame 4400~\AA\ apparent AB magnitude, $m_{\rm AB}(4400)$, from
our broad-band imaging.  Magnitudes in the AB system \markcite{Oke:74}(Oke 1974) are
defined by $m_{\rm AB}(\lambda) \equiv -2.5 \log f_\nu(\lambda) +
23.90$, where $f_\nu$ is measured in $\mu$Jy.  For the quasar CXO50,
our highest-redshift source, we do not have an imaging data point
long-ward of rest-frame 4400~\AA:  for this source we instead assume a
standard quasar optical spectral index, $f_\nu \propto \nu^{-0.5}$
\markcite{Richstone:80, Schneider:92}(\eg Richstone \& Schmidt 1980; Schneider {et~al.} 1992) and interpolate the flux
density from the $I$-band magnitude.  A cosmology-dependent luminosity
distance is used to relate apparent magnitude $m_{\rm AB}$ to absolute
magnitude $M_{\rm AB}$, and the same optical spectral index is used to
calculate the offset between Vega-based $M_B$ and AB-system $M_{\rm
AB}(4400)$: $M_B = M_{\rm AB}(4400) + 0.12$
\markcite{Kennefick:95}(\eg Kennefick, Djorgovski, \&  de~Calvalho 1995).  As seen in Table~3, the
differences in absolute magnitude are not large between the two
cosmologies considered, $M_B^\Lambda = M_B^{\rm EdS} \pm 0.3$.  For the
figures below we plot only the Einstein-de~Sitter values.

For $\Omega = 1$ and $\Lambda = 0$, the luminosity in the rest frame
energy band $E_1 - E_2$, $L_{E_1 - E_2}$, is then related to the flux
observed in that same energy band, $S_{E_1 - E_2}$, by $$L_{E_1 - E_2}
= 4 \pi \left( {{2 c} \over H_0} \right)^2 {{ (1 + z - \sqrt{1+z})^2}
\over {(1+z)^{2 - \Gamma}}} S_{E_1 - E_2}$$ \markcite{Hogg:99}(\eg see Hogg 1999),
where we adopt an X-ray spectral index $\Gamma = 1.4$ for our sample.
Fig.~\ref{fig_Xlum} plots the X-ray luminosity of our spectroscopic
sample against the rest-frame $B$-band luminosity, $M_B$.
Fig.~\ref{fig_zlum} plots X-ray luminosity against redshift for our
sample, fortified with the HDF-N {\it Chandra} sources from
\markcite{Hornschemeier:01}Hornschemeier {et~al.} (2001).  We have recalculated their luminosities for
our cosmology and $\Gamma = 1.4$.

We have classified the optical spectra in three broad categories:  AGN,
galaxies, and stars, following a scheme similar to that of
\markcite{Schmidt:98}Schmidt {et~al.} (1998) and \markcite{Hornschemeier:01}Hornschemeier {et~al.} (2001).  A description of the
categories follows:

\begin{itemize}

\item{\bf AGN:}   We classify sources as AGN that have either (1)
broadened emission lines (FWHM $\simgt 1000 \kms$) or (2)
high-ionization state emission lines such as [\ion{Ne}{5}].
Unfortunately the coarse spectral resolution at which many of these data
were taken hinders us from separating broad-lined AGN from narrow-lined
AGN.  Seven of the eighteen sources in the current sample are
classified as AGN.  They are predominately identified with the most
X-ray-luminous sources, $L_X \simgt 10^{43} \ergs$.  Many are also
luminous at optical wavelengths (Fig.~\ref{fig_Xlum}).

\item{\bf Emission-Line and Early-Type Galaxies:} Extragalactic sources
without obvious AGN features in their optical spectra are classified as
galaxies, though this does not rule out the presence of an active
nucleus.  We distinguish between galaxies showing emission lines and
early-type galaxies, where the latter have redder continua marked by
continuum breaks at 2640 \AA\ (B2640), 2900 \AA\ (B2900), and 4000
\AA\ (D4000).  Many of the emission-line galaxies show early-type
features as well.  Nine of the eighteen sources in the current sample
are classified as galaxies.  For the lower X-ray luminosity sources,
stellar processes such as binaries, winds, and supernovae can produce
the X-ray emission.  For the higher X-ray luminosity sources, a buried
active nucleus is likely present (see \S 6.4).

\item{\bf Stars:}  Two of the eighteen sources in the current sample
are classified as low-mass Galactic dwarfs.  \markcite{Hornschemeier:01}Hornschemeier {et~al.} (2001)
also report two X-ray-emitting M4 dwarfs in the {\it Chandra} field
encompassing the HDF-N.  

\end{itemize}

\subsection{Notes on Individual Sources}

{\bf CXO39 ({\boldmath $z=0.573$}, AGN):}  This X-ray-luminous
($L_{0.5-2} = 10^{43.7}~ \ergs$, $L_{2-10} = 10^{43.8}~ \ergs$) source
shows many high equivalent width emission features with deconvolved
velocity widths FWHM $\approx 2000 \kms$.  High-ionization species of
Ne are also detected.  The spectra were taken without an order-blocking
filter, making the continuum unreliable long-ward of 7000 \AA.

{\bf CXO40 ({\boldmath $z=0.622$}, emission-line galaxy):}  This galaxy
shows a spectrum typical of young, star-forming galaxies:  emission
from \oii, the \oiii\ doublet, and absorptions from both CaH$+$K and
the hydrogen Balmer lines.  The X-ray luminosity is approximately an
order of magnitude lower than that typical of sources with obvious AGN
features in their optical spectra ($L_{0.5-2} = 10^{42.2} \ergs,
L_{2-10} = 10^{42.4} \ergs$), though this does not rule out some
fraction of the X-ray emission deriving from a nuclear supermassive
black hole.

{\bf CXO41 ({\boldmath $z=1.329$}, AGN):}  This source shows broad
\mgii\ (FWHM $\sim 2500$ \kms) and \oii\ emission.  The optical
spectrum long-ward of 7000 \AA\ is suggestive of a normal star-forming
galaxy, a higher redshift analog of CXO40.  However, the broad \mgii,
high X-ray luminosity ($L_{2-10} = 10^{43.6} \ergs$), and strong core in
the {\em HST} image conclusively show that an AGN plays a prominent
role in this galaxy.

{\bf CXO42 ({\boldmath $z=1.035$}, AGN):}  This source shows broad
\mgii\ emission (FWHM $\approx 5300$ \kms) and narrow emission lines
from high-ionization species of Ne.  The \nev\ lines are individually
unresolved at our coarse spectral resolution (FWHM $\simlt 1300$
\kms).  No stellar features are seen.  We classify the source as an
AGN.  The {\it HST} image shows the source to be unresolved.

{\bf CXO44 ({\boldmath $z=0.725$}, emission-line galaxy):}  The optical
spectrum of this source shows weak, spectrally unresolved
\oii\ emission superposed on an early-type galaxy spectrum.  No obvious
AGN features are evident in the optical spectrum.  The galaxy is
moderately X-ray luminous ($L_{0.5-2} = 10^{42.0} \ergs, L_{2-10} =
10^{42.9} \ergs$).  The {\it HST} image shows a complicated, merging
system, suggestive of merger-induced (nuclear?) activity powering the
X-ray emission in this system.

{\bf CXO49 ({\boldmath $z=1.017$}, emission-line galaxy):}  This source
shows a single emission line straddled by a faint, blue continuum.  The
most likely identification of such a line is
\oii\ \markcite{Stern:00d}(\eg Stern {et~al.} 2000).  The morphology from {\it HST} image
shows an unresolved source.

{\bf CXO50 ({\boldmath $z=3.093$}, AGN):}  This broad-lined AGN is the
highest-redshift source in the current sample.  It shows the classic
features of a broad absorption line quasar (BALQSO).  Note that the
absorption blue-ward of \ciii\ is telluric in origin, from the
atmospheric A-band.

{\bf CXO54 ({\boldmath $z=0.569$}, AGN):}  This source is similar to
CXO41 and CXO42, showing broad \mgii\ (FWHM $\sim 6300$ \kms) and
narrow \oii\ emission.  The [\ion{O}{2}] strength is considerably
weaker in this source compared to CXO41.  We also detect
\neiii\ emission from this source.  The redshift places this source at
the same redshift of the $z=0.570$ cluster.

{\bf CXO57 ({\boldmath $z=1.194$}, AGN):}  Similar to many of the other
galaxies, this source shows broad \mgii\ (FWHM $\sim 6600$ \kms) and
strong, narrow \oii\ emission.

{\bf CXO63 ({\boldmath $z=0.899$}, AGN):}  This source shows strong
\mgii\ and \hbeta\ emission of moderate width, FWHM $\approx 2000$
\kms.  Emission from \cii\ and \neiv\ are also detected.  Again, the
spectra were taken without an order-blocking filter, making the
continuum unreliable long-ward of 7000 \AA.

{\bf CXO65 ({\boldmath $z=0.747$}, emission-line galaxy):}  This source
shows an unremarkable galaxy spectrum with a prominent 4000 \AA\ break
(D4000) and unresolved \oii\ emission.  No \mgii\ emission is
detected.  The source is moderately X-ray luminous, $L_{0.5-2} =
10^{42.0} \ergs$ and $L_{2-10} = 10^{42.9} \ergs$.

{\bf CXO66 ({\boldmath $z=0.336$}, emission-line galaxy):}  This is the
least X-ray luminous galaxy in the current spectroscopic sample
($L_{0.5-2} = 10^{41.1} \ergs, L_{2-10} = 10^{42.0} \ergs$).  The
spectrum shows classic early-type galaxy features (\ie Ca H$+$K
absorption, D4000, H$\delta$ absorption) with weak \oii\ and
\oiii\ emission evident.

{\bf CXO72 (M7 V):}  CXO72 has spectral characteristics intermediate
between those of late type dwarfs and giants, suggestive of an
intermediate gravity object.  The spectral energy distribution between
6500 and 9400 \AA\ and lack of CaH absorption near 6950 \AA\ are
suggestive of an M7~III giant, while the strengths of the \ion{K}{1}
doublet near 7700 \AA\ and the \ion{Na}{1} feature near 8400 \AA\ are
more suggestive of an M7~V dwarf.  We favor the latter interpretation
as it provides a more realistic distance for a low-extinction, high
Galactic latitude source:  700~pc, using the \markcite{Kirkpatrick:94}Kirkpatrick \& McCarthy (1994)
optical absolute magnitude tabulation for late-type dwarfs.  Unlike the
other Galactic dwarf identified in this paper or the two X-ray-emitting
dwarfs discussed in \markcite{Hornschemeier:01}Hornschemeier {et~al.} (2001), we detect hard X-rays
from CXO72.  As seen in Fig.~\ref{fig_Xopt}, the soft X-ray-to-optical
flux ratio is also quite different than the other X-ray dwarfs.
Assuming the dwarf interpretation, the corresponding hard X-ray
luminosity is $2 \times 10^{29} \ergs$.  We discuss this source further
in \S 6.3.

{\bf CXO128 ({\boldmath $z=0.542$}, early-type galaxy):}  This galaxy,
detected only in the soft X-ray band, has an early-type spectrum
lacking any emission features.  The Ca H+K absorption doublet, 4000
\AA\ break (D4000), and G-band are very prominent.  The $I$-band
magnitude is approximately $L^*$ for a $z\sim 0.55$ giant elliptical
galaxy.  The redshift of this source is similar to the group at $z =
0.543$, slightly foreground and south of the $z = 0.570$ cluster
\markcite{Holden:01}(see Holden {et~al.} 2001).

{\bf CXO138 ({\boldmath $z=1.26$}, early-type galaxy):}  This extremely
red galaxy ($R-K = 6.05$, $K=17.90$) has a spectrum similar to that of
the old, dead, and red radio galaxy LBDS53W091 \markcite{Dunlop:96,
Spinrad:97}(Dunlop {et~al.} 1996; Spinrad {et~al.} 1997 -- shown for
comparison in Fig.~11):  the optical spectrum is devoid of emission
lines and shows continuum breaks at rest-frame 2640 \AA\ (B2640), 2900
\AA\ (B2900), and 4000 \AA\ (D4000).  These features are seen in {\em
International Ultraviolet Explorer} ({\em IUE}) spectra of F-type
main-sequence stars \markcite{Fanelli:92}(Fanelli {et~al.} 1992),
implying this is an old ($\sim 3$~Gyr) galaxy at high-redshift.  The
redshift matches that of the two high-redshift clusters targeted by the
{\em Chandra} observations.  The source, detected in the soft X-ray
band near our detection limit, has an X-ray luminosity higher than the
other spectroscopically non-active galaxies.

{\bf CXO139 (M4 V):}  This M4 dwarf is detected only in the soft band,
similar but slightly fainter than the two M4 dwarfs detected in {\em
Chandra} observations of the field encompassing the HDF-N
\markcite{Hornschemeier:01}(Hornschemeier {et~al.} 2001).  The X-ray-to-optical flux ratio is similar
to those stars.  Using optical/near-infrared absolute magnitudes of an
M4~V dwarf calculated by \markcite{Kirkpatrick:94}Kirkpatrick \& McCarthy (1994), we derive a distance
of $\approx 420$~pc for CXO139.  The corresponding X-ray luminosity is
$L_{0.5-2} = 6.4 \times 10^{27} \ergs$.  Again, this luminosity is
similar to the two soft X-ray-emitting M4 dwarfs reported in
\markcite{Hornschemeier:01}Hornschemeier {et~al.} (2001).  We discuss X-ray emission from Galactic
sources further in \S 6.3.

{\bf CXO164 ({\boldmath $z=0.750$}, early-type galaxy):}  The spectrum
of this source is devoid of emission lines and shows absorption from Ca
H$+$K, a D4000 break, and G-band absorption.  The source is detected
only in the hard X-ray band, suggesting that the source may harbor an
obscured AGN from which no feature is evident in the optical data.

{\bf CXO167 ({\boldmath $z=0.432$}, early-type galaxy):}  Similar to
CXO164, CXO167 is detected only in the hard X-ray band and shows an
early-type galaxy spectrum in the optical devoid of emission lines.
The feature near the redshifted \oiii\ line is spurious, the residual
from a cosmic ray.

\subsection{Galactic X-ray sources}

Two of the sources in our sample are spectroscopically identified with
late-type Galactic dwarfs.  \markcite{Hornschemeier:01}Hornschemeier {et~al.} (2001) also find two M4
dwarfs in their survey of the field encompassing the HDF-N.  Three of
these four sources are very similar:  mid-M dwarfs detected only in the
soft X-ray band with $\log (f_X / f_R) \approx -2$.  The remaining
source, CXO72, is a late-M dwarf, detected out to 7~keV and has $\log
(f_X / f_R) \approx 0.5 - 1$.  We briefly review X-ray emission from
late-type stars followed by a  discussion of this intriguing new
source.

Phenomenologically, the fraction of stars showing stellar activity is
found to increase with lower stellar mass, reaching 100\%\ at spectral
type M7, then to decrease such that few objects later than type L5 show
activity \markcite{Hawley:96, Gizis:00}(Hawley, Gizis, \& Reid 1996; Gizis {et~al.} 2000).  Physically, the observed
activity, seen in the form of elevated \halpha\ and/or X-ray emission,
is believed to result from collisional heating of ions and electrons
along magnetic field lines in stellar chromospheres.  For typical
chromosphere temperatures, such stars are only detected in the soft
X-ray band.  Stars more massive than $\simeq 0.3 M_\odot$ have both
convective and radiative zones which allow the formation of a stable
internal dynamo through the $\alpha-\Omega$ mechanism
\markcite{Parker:55}(Parker 1955).  More rapidly spinning stars have stronger magnetic
fields and thus show enhanced activity.  Angular momentum loss from
stellar winds causes stars to spin-down as they age, with the more
massive stars exhibiting stronger winds and thus spinning down more
rapidly.  The rise in stellar activity from spectral type K5 to
spectral type $\approx$~M9 thus simply tracks the fraction of stars
sufficiently young for their given spectral type and associated
spin-down rate that the internal dynamo can power stellar activity.
For spectral types M7 -- M9, the spin-down rate is apparently longer
than the Hubble time.  This scenario is supported by the observed
correlation between activity and rotation rate in late type stars
\markcite{Kraft:67, Basri:87}(\eg Kraft 1967; Basri 1987).

%

The soft-X-ray emitting M4 dwarfs detected by {\it Chandra} in the deep
surveys is consistent with this scenario. Curiously, however, none of
these sources show elevated H$\alpha$ emission.  CXO72 is a more
unusual source.  As seen in Fig.~\ref{fig_Xopt}, for its optical
$R$-band magnitude, this star is much brighter in the soft X-ray than
the other Galactic X-ray sources discussed here.  Some of the
difference is likely due to the redder spectral energy distribution of
its later spectral class.  However, CXO72 is also detected in the hard
X-ray band implying a more energetic energy production mechanism than
in the soft-band-only Galactic sources.  \markcite{Rutledge:00}Rutledge {et~al.} (2000) report the
{\it Chandra} detection of an X-ray flare from the lithium-bearing M9
brown dwarf LP944$-$40, with {\em soft} X-ray emission detected for
1$-$2~hr during a 12.1~hr observation.  Unlike that source, the X-ray
emission from CXO72 is non-transient (Fig.~\ref{fig_lightcurve}).  We
posit that CXO72 is most likely associated with a very low mass binary
system, among the lowest mass such systems observed, with accretion
powering the X-ray emission.  \markcite{Burgasser:00}Burgasser {et~al.} (2000) recently reported
strong H$\alpha$ emission from a T dwarf (methane brown dwarf)
discovered in the 2MASS survey.  They suggest a close [$a \sim (4-20)
R_J$] interacting binary system with Roche lobe overflow powering the
observed emission.  Both the H$\alpha$ emission in the T~dwarf system
and the X-ray emission in CXO72 are not seen to temporally vary in
brightness, again suggesting that flaring is not the source of observed
stellar activity.  Alternatively, most of the hard X-rays from CXO72
come from the second observing period (see Fig.~\ref{fig_lightcurve}),
suggesting that time-variable phenomena could be involved.

\subsection{X-Ray Emission from Normal Galaxies}

Nine of the X-ray sources discussed here show optical spectra devoid of
strong or obvious AGN features.  As seen in Fig.~\ref{fig_zlum}, these
sources tend to lie at low redshift ($z \simlt 1$) with $L_{0.5-2}
\simlt 10^{42} \ergs$ and $L_{2-10} \simlt 10^{43} \ergs$.  Many of
the sources are detected in only one X-ray band.  Similar sources are
reported in \markcite{Mushotzky:00}Mushotzky {et~al.} (2000), \markcite{Barger:01}Barger {et~al.} (2001),
\markcite{Hornschemeier:01}Hornschemeier {et~al.} (2001), and \markcite{Tozzi:01}Tozzi {et~al.} (2001).  We briefly review
physical processes which might produce such emission.

\subsubsection{Actively Star-Forming Galaxies} 

With a typical $L_X / L_{\rm bol} \sim 10^{-7}$ thought to derive from
shocks developing in unsteady wind outflows \markcite{Pallavicini:87}(Pallavicini {et~al.} 1987),
isolated late-type stars are not prodigious X-ray emitters.  However, a
galaxy with a significant young stellar population will produce soft
and hard X-rays from several astrophysical mechanisms, primarily
associated with the beginning and end stages of massive star evolution
\markcite{Helfand:01}(\eg see Helfand \& Moran 2001).  Pre-main sequence stars will emit
X-rays in their T~Tauri phase \markcite{Koyama:96}(\eg Koyama {et~al.} 1996).  Shocks and hot
gas associated with stellar winds, galactic winds, and supernovae will
also produce X-ray emission, though the dominant source of hard X-ray
emission in a starburst galaxy will be accretion-driven from high-mass
X-ray binaries (HMXBs).  

From a sample of major Local Group Galaxies and Galactic OB stars within
3~kpc of the Sun, \markcite{Helfand:01}Helfand \& Moran (2001) infer a specific $2-10$~keV X-ray
luminosity per O~star of $2 - 20 \times 10^{34} \ergs$.  Using the models
of \markcite{Leitherer:95}Leitherer, Carmelle, \&  Heckman (1995) we can then relate hard X-ray luminosity to the
galaxy star formation rate, $SFR$.  For a Salpeter initial mass function
(IMF) with slope 2.35, an upper mass cutoff 100 $M_\odot$, and solar
metallicity, the models of \markcite{Leitherer:95}Leitherer {et~al.} (1995) show that a region
producing $1 M_\odot~ {\rm yr}^{-1}$ for at least $10^7$~yr will have
$2.5 \times 10^4$ O~stars.  Therefore, $$SFR = 2-20 \times 10^{-40}~
L_{2-10}~ M_\odot~ {\rm yr}^{-1},$$ where the hard X-ray luminosity
$L_{2-10}$ is measured in \ergs.

All of the galaxies in our sample showing evidence of star formation
(\ie sources classified as emission-line galaxies) are indeed detected
in the hard X-ray band, with $L_{2-10} \approx 10^{42} - 10^{43}
\ergs$, implying a wide range of possible star formation rates.
However, \oii\ luminosity, $L_{[OII]}$, also provides a crude estimate
of star formation rate \markcite{Kennicutt:92}(Kennicutt 1992):  $$SFR \approx 5 \times
10^{-41}~ L_{[OII]}~ M_\odot~ {\rm yr}^{-1},$$ where $L_{[OII]}$ is
measured in units of \ergs.  For our sample of emission-line galaxies,
the SFR derived from the X-ray luminosity is typically more than three
orders of magnitude greater than that derived from the
\oii\ luminosity.  The implication is that a buried AGN is likely
producing much of the X-ray emission (see \S 6.4.3) though
dust-enshrouded star formation cannot be ruled out.

\subsubsection{Early-Type Galaxies}

The {\it Einstein Observatory} found that some early-type galaxies are
powerful X-ray emitters, with X-ray luminosities correlated with
optical luminosities \markcite{Forman:85}(Forman, Jones, \& Tucker
1985), albeit it with a significant dispersion \markcite{Brown:98}($L_X
\propto L_B^{1.7-3.0}$; \eg Brown \& Bregman 1998).  For the X-ray
luminous galaxies, hot ($\sim 10^7$~K) interstellar gas lost by stars
during normal stellar evolution is thought responsible for producing
the X-ray emission.  The gas thermalizes in the galactic potential well
and then cools as it flows to the center of the galaxy.  The X-ray
spectra of these galaxies are dominated by thermal emission at $kT \sim
0.8$~keV.  It is thought that the X-ray-faint galaxies have lost much
of their interstellar gas through galactic winds or from ram pressure
stripping by ambient intercluster or intergroup gas.  These X-ray-faint
galaxies show a harder X-ray spectral component, $\sim 5-10$~keV
\markcite{Matsumoto:97}(Matsumoto {et~al.} 1997), with a very soft
($\sim 0.2$~keV) component.  Recent {\it Chandra} observations of the
X-ray-faint elliptical galaxy NGC~4697 resolves most of the soft and
hard X-ray emission into point sources, the majority of which are
low-mass X-ray binaries \markcite{Sarazin:00}(LMXBs; Sarazin, Irwin, \&
Bregman 2000).

In Fig.~\ref{fig_Xlum} we plot rest-frame $0.5-2$~keV luminosity,
$L_{0.5-2}$, against rest-frame $B$-band absolute magnitude, $M_B$, for
the spectroscopically-observed {\it Chandra} sources from our survey.
We also plot 34 early-type galaxies observed by {\it ROSAT} by
\markcite{Brown:98}Brown \& Bregman (1998).  Their sample comprises the 34 optically-brightest,
early-type galaxies from the optically-selected, flux-limited sample of
\markcite{Faber:89}Faber {et~al.} (1989), excluding sources with $\vert b \vert < 20\deg$,
dwarfs, and M87, known to contain an X-ray bright AGN.  Only two
sources in our sample classified as early-type galaxies --- CXO128 and
CXO138 --- are detected in the soft X-ray band.  Both reside within
portions of the $L_{0.5-2} - M_B$ plot (Fig.~\ref{fig_Xlum}) consistent
with the early-type galaxies studied by \markcite{Brown:98}Brown \& Bregman (1998).  Several of
the sources classified as having emission-line galaxy spectra also have
soft X-ray luminosities consistent with the bright early-type sample.
The soft X-ray luminosities for these sources can presumably be
explained by hot gas and stellar sources.  On the other hand, sources
CXO164 and CXO167 are both classified as early-type galaxies from their
optical spectra but are detected only in the {\em hard} X-ray band.  The
X-ray emission in these two sources presumably is powered by a buried
AGN.  Interestingly, the two soft-band-only early-type galaxies are at
redshifts where the Lynx field has known clusters while the two
hard-band-only early-type galaxies are at redshifts lacking large-scale
structure in the Lynx field.

\subsubsection{Obscured AGN}

Unification models of AGN invoke orientation as one of the dominant
parameters that affect the observed optical properties of an active
nucleus.  For each broad-lined quasar whose optical emission is
Doppler-boosted along our line of sight, a population of {\em
misdirected} quasars should exist whose AGN optical emission is
severely attenuated.  \markcite{Norman:01}Norman {et~al.} (2001) report a clear example of one
such source, where the optical spectrum shows strong, narrow-lined
emission.  The well-studied high-redshift radio galaxies (HzRGs) are
the radio-loud component of this population:  they also typically only
exhibit narrow emission lines \markcite{McCarthy:93, Stern:99a}(\eg McCarthy 1993; Stern {et~al.} 1999).
Presumably the more centrally-condensed broad-lined region (BLR) of
these sources has been obscured while the spatially-extended
narrow-lined region (NLR) has not been.  These spatial scales are
consistent with reverberation mapping studies of AGN.  For a less
luminous active nucleus, the stellar emission from the galaxy might
dominate the optical spectral energy distribution, providing an optical
spectrum devoid of AGN features.  The effects of obscuration decrease
at high energies, allowing X-ray detection of obscured AGN in
apparently normal galaxies.

\subsection{Comparison to X-Ray Quiet Galaxies}

How do the X-ray-emitting galaxies compare to other galaxy
populations?  In Fig.~\ref{fig_kz} we show the $K$-band brightness of
sources from our survey plotted against spectroscopic redshift.
High-redshift radio galaxies from several radio surveys are also
plotted \markcite{DeBreuck:01}(see {De~Breuck} {et~al.} 2001, and
references therein).  At each redshift, HzRGs are the most luminous
galaxies known at observed 2$\mu$m \markcite{DeBreuck:01}(\eg see
Fig.~10 in {De~Breuck} {et~al.} 2001).  Since 2$\mu$m samples stellar
emission from the low-mass stars which dominate the baryon content of a
galaxy, this is interpreted as HzRGs being the most massive systems at
each cosmic epoch \markcite{Dey:99c}(\eg Dey 1999, and references
therein).  In Fig.~\ref{fig_kz} we also plot 219 redshifts from our
followup of $K_s < 20$ sources in the Lynx field of SPICES.  As we can
see from this figure, the {\it Chandra} sources are among the brighter
2$\mu$m sources at each redshift, though they are not as luminous as
the HzRGs identified from larger area surveys.

\section{Conclusions}

We present first results from our deep, 184.7~ks {\it Chandra} map of
the Lynx field.  This is the third-deepest X-ray field observed thus
far by {\it Chandra}, exceeded in depth only by the megasecond
campaigns in the HDF-N and CDF-S \markcite{Hornschemeier:01,
Tozzi:01}(Hornschemeier {et~al.} 2001; Tozzi {et~al.} 2001).  We
present a catalog of all 153 unresolved field X-ray sources detected in
this survey which will provide essential information on the sources
responsible for the soft ($0.5-2$~keV) and hard ($2-10$~keV) X-ray
backgrounds.  We discuss basic results from the X-ray data alone,
showing the fainter X-ray sources predominantly have steeper X-ray
spectra.  We also present multiband optical and near-IR ($BRIzJK_s$)
identifications for the $\simeq 66$\%\ of Lynx {\it Chandra} sources
which are within the area studied by SPICES, the $K$-selected imaging
and redshift survey we are pursuing \markcite{Eisenhardt:01}(see
Eisenhardt {et~al.} 2001).  We find $\sim 40$\%\ of the
spectroscopically-studied subsample are identified with galaxies
showing obvious AGN features (\ie broad lines and high-ionization
emission), $\sim 50$\%\ are identified with galaxies showing no obvious
AGN features, and 2 ($\sim 10$\%) of the sources are identified with
late-type Galactic dwarfs.  One Galactic source, CXO72, is unusual,
having an M7 spectral classification and an unusually hard X-ray
spectrum.  We discuss this sparse spectroscopic sample in relation to
other recently reported deep {\it Chandra} fields and briefly review
X-ray emission from sources not showing obvious AGN features in their
optical spectra.  We conclude indirectly that many such sources in our
survey must indeed harbor an obscured AGN.  The catalogs presented here
will prove useful for more complete spectroscopic followup allowing
direct comparison to models of the X-ray background and studies of the
fraction of massive galaxies harboring active, enshrouded nuclei.

\acknowledgments

We are indebted to numerous colleagues for stimulating conversation and
insight regarding this work.  In particular, we thank John Gizis, Davy
Kirkpatrick, John Stauffer, Jim Liebert, and Bob Rutledge for their
input regarding CXO72, the hard X-ray-emitting, late-type Galactic
dwarf.  We also thank Fiona Harrison, Peter Mao, David Helfand, and Ed
Moran for contributing discussion regarding the whole breadth of this
project.  We thank Roberto della~Ceca for providing us with the
hard-band $\log N - \log S$ results from {\it ASCA}.  The authors wish
to extend special thanks to those of Hawaiian ancestry on whose sacred
mountain we are privileged to be guests.  Without their generous
hospitality, many of the observations presented herein would not have
been possible.  The work of DS and PE were carried out at the Jet
Propulsion Laboratory, California Institute of Technology, under a
contract with NASA.  The work of SD was supported by IGPP/LLNL
University Collaborative Research Program grant \#02$-$AP$-$015, and
was performed under the auspices of the U.S. Department of Energy by
University of California Lawrence Livermore National Laboratory under
contract No.\ W$-$7405$-$Eng$-$48.  AD's research is supported by NOAO
which is operated by AURA under cooperative agreement with the NSF.
This work has been supported by the following grants:  NSF grant
AST~95$-$28536 (HS) and NSF CAREER grant AST~9875448 (RE).

\eject

\eject


\begin{figure}[!t]
\begin{center}
\end{center}

\caption{False-color X-ray image of the Lynx field, composed from the
{\it Chandra} data.  North is to the top, east is to the left.
Red/green/blue maps were composed using the following energy channels:
0.3$-$1~keV, 1$-$2~keV, and 2$-$7~keV.  Bluer sources have harder
spectral indices.  Red circles identify the three diffuse X-ray sources
associated with hot gas in distant galaxy clusters.  The northern
cluster is at $z = 0.57$ (see Holden \etal 2001) while the flanking,
southern clusters are at $z \simeq 1.27$ (see Stanford \etal 2001).
The outer, white square indicates the area of our deep, optical $BRIz$
images and the inner, white square indicates the area of our deep
near-infrared $JK_s$ images.}

\label{Xcolorfig}
\end{figure}


\begin{figure}[!t]
\begin{center}
\plotfiddle{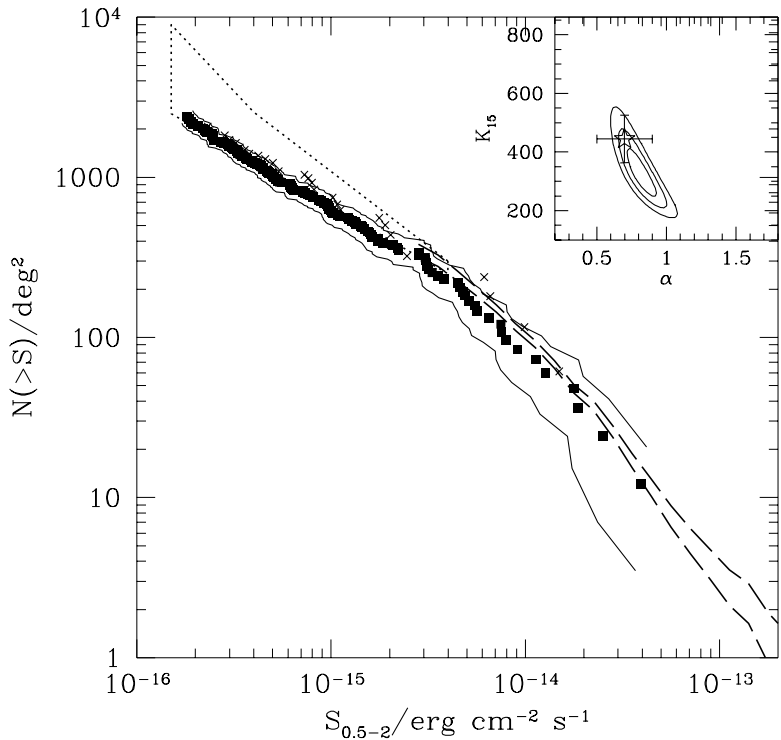}{3.3in}{0}{60}{60}{-200}{-125}
\end{center}

\caption{The $\log N - \log S$ relation in the soft X-ray band from the
deep {\it Chandra} observations of the Lynx field (filled squares).
Crosses are from the 101~ks {\it Chandra}/ACIS-S3 observations of the
SSA13 (Mushotzky \etal 2000).  Dashed lines are {\it ROSAT} counts from
the Lockman Hole (Hasinger \etal 1998) and dotted contour is the
extrapolation from fluctuation analysis of {\it ROSAT} data (Hasinger
\etal 1993).  Insert shows the maximum-likelihood fit to the parameters
in the  $\log N - \log S$ relation for $N(>S_{0.5-2}) = K_{15}
(S_{0.5-2} /2 \times 10^{-15})^{- \alpha}$.  The contours correspond to
$1 \sigma$, $2 \sigma$, and $3 \sigma$.  The star is the fit from
Mushotzky \etal (2000) at $S_{0.5-2} = 2 \times 10^{-15} \ergcm2s$,
with an error bar corresponding to their 68\%\ confidence limit.}

\label{fig_NSsoftXRB}
\end{figure}


\begin{figure}[!t]
\begin{center}
\plotfiddle{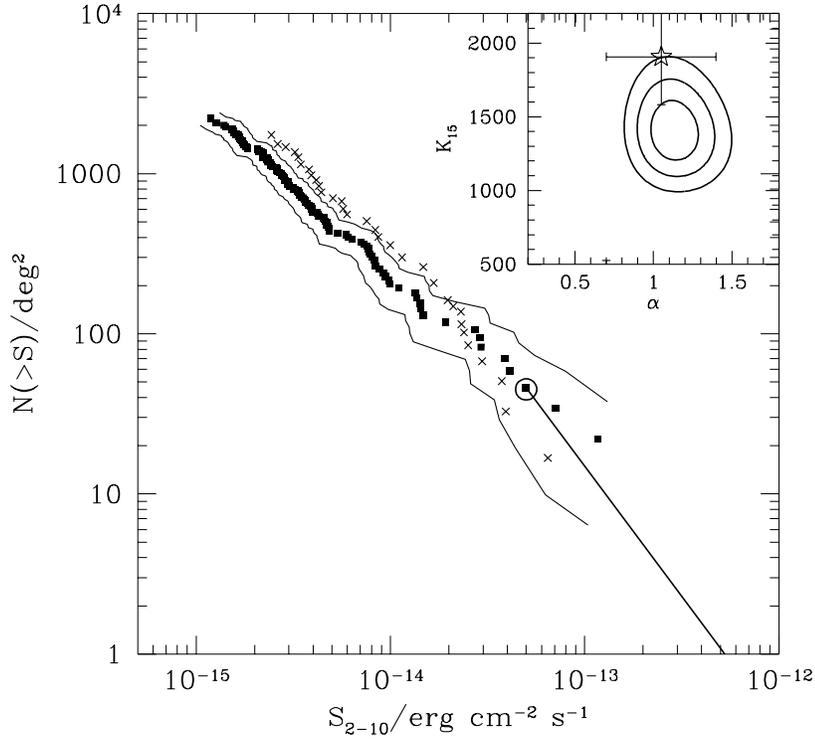}{3.3in}{0}{60}{60}{-200}{-125}
\end{center}

\caption{The $\log N - \log S$ relation in the hard X-ray band from the
deep {\it Chandra} observations of the Lynx field (filled squares).
Crosses are from the 101~ks {\it Chandra}/ACIS-S3 observations of the
SSA13 (Mushotzky \etal 2000).  The large open circle at high flux is
from {\it ASCA} and {\it BeppoSAX} (Giommi \etal 1998, Ueda \etal 1999)
and the continuous line is the fit to the {\it ASCA} counts in the
range $(1 - 10) \times 10^{-13}$ \ergcm2s (della Ceca \etal 1999).
Upper and lower solid lines indicate uncertainties due to Poisson noise
(1$\sigma$), calculated for $\Gamma = 1.4$.  The insert show the 1,2,
and 3$\sigma$ maximum likelihood fits to the relation $N(>S_{2-10}) =
K_{15}(S_{2-10}/2 \times 10^{-15})^{-\alpha}$.  The star shows the
results of Mushotzky \etal (2000) at $S = 2 \times 10^{-15} \ergcm2s$,
with their 68\%\ confidence limit also illustrated.}

\label{fig_NShardXRB}
\end{figure}


\begin{figure}[!t]
\begin{center}
\plotfiddle{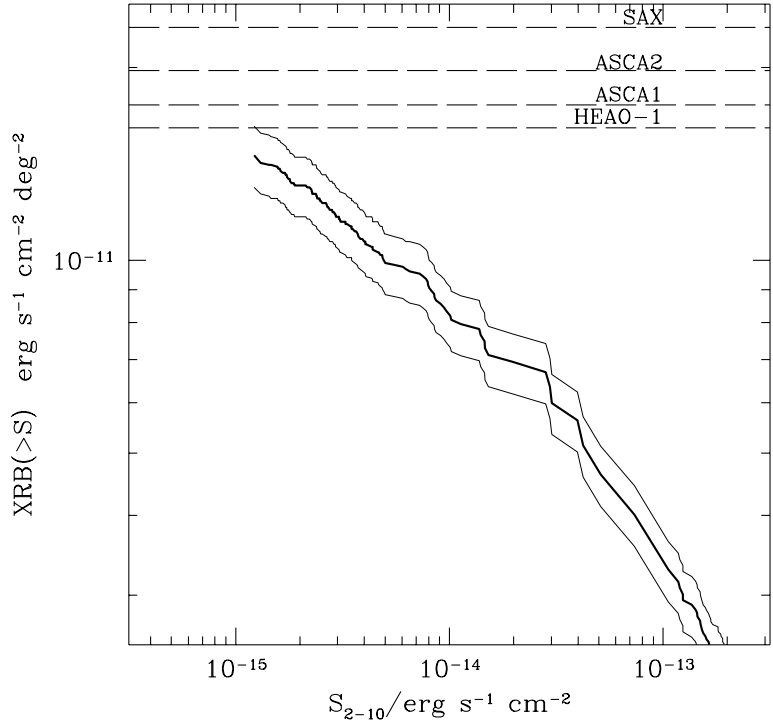}{3.3in}{0}{60}{60}{-200}{-125}
\end{center}

\caption{Contribution to the hard X-ray background as a function of the
flux of resolved sources.  For $S_{2-10} < 10^{-13}$ \ergcm2s\ we show
results from the present work.  For $S_{2-10} > 10^{-13}$ \ergcm2s\ we
include the contribution derived from the {\em ASCA} sample of
della~Ceca \etal (1999).  The horizontal lines in the upper part of the
plot refer to previous measurements of the hard X-ray background.  From
bottom to top, they are:  Marshall \etal (1980; {\it HEAO-1}), Ueda
\etal (1999; {\it ASCA1}), Ishisaki \etal (1999; {\it ASCA2}), and
Vecchi \etal (1999; {\it BeppoSAX}).}

\label{fig_hardXRB2}
\end{figure}


\begin{figure}[!t]
\begin{center}
\plotfiddle{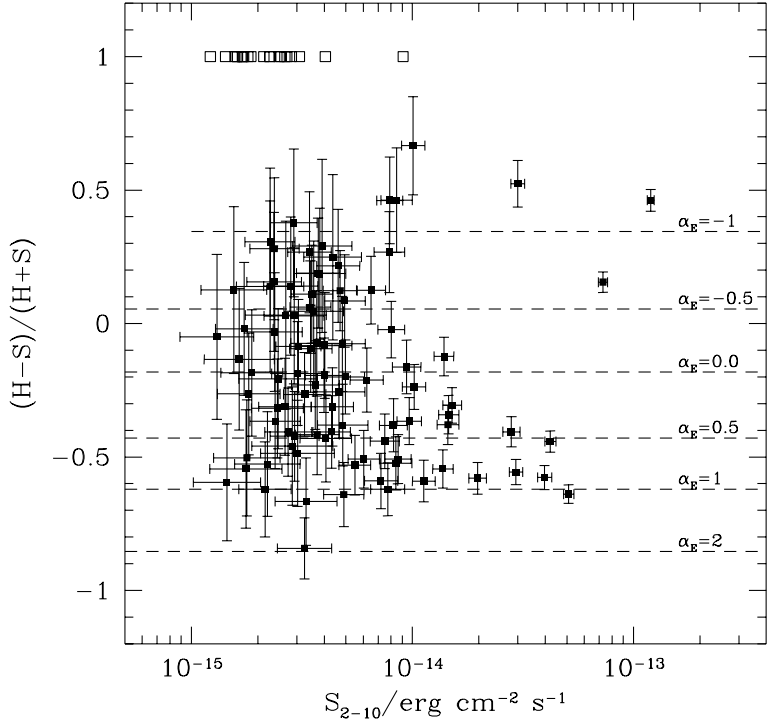}{3.3in}{0}{60}{60}{-200}{-125}
\end{center}

\caption{Hardness ratio $HR \equiv (H-S/H+S)$ of X-ray sources as a
function of hard X-ray flux.  Sources detected only in the hard X-ray
band are shown at a hardness ratio of $1$, while sources detected only
in the soft X-ray band are not shown.  Dashed horizontal lines are
power-law models with different energy index ($\alpha_E \equiv \Gamma -
1$) computed assuming the Galactic value $N_H \simeq 2 \times 10^{20}
\cm2$ and convolved with a mean ACIS response matrix at the aimpoint.
The three bright hard X-ray sources with large values of their hardness
ratio are, in order of decreasing hard X-ray flux, CXO12 ($HR$ = 0.46),
CXO37 ($HR$ = 0.16), and CXO36 ($HR$ = 0.53).  Ohta \etal (1996) report
that CXO12 is a Type~2 quasar at $z = 0.9$.}

\label{fig_softhard}
\end{figure}


\begin{figure}[!t]
\begin{center}
\plotfiddle{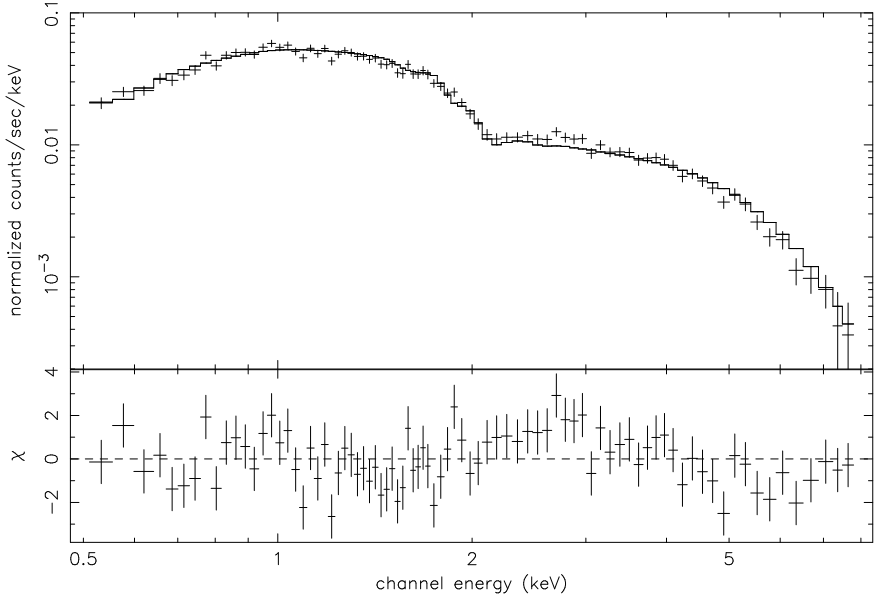}{3.3in}{-90}{50}{50}{-220}{325}
\end{center}

\caption{Stacked spectrum of the total sample of sources in the Lynx
field fitted in the energy band $0.5-8$ keV with an absorbed power law
with variable column density.    The best fit slope is $\Gamma = 1.36
\pm 0.04$ and the best fit absorbing column is $N_H = (2.3 \pm 1.4)
\times 10^{20}$ cm$^{-2}$, consistent with the Galactic value (errors
at 90\% confidence level).   The model, with a $\chi_\nu^2 = 1.16$, is
in agreement with the average slope of the unresolved hard X-ray
background $\langle \Gamma \rangle = 1.4$.  The solid line is the
best-fit model, while the lower panel shows the standard deviations in
each energy bin.}

\label{fig_allsources}
\end{figure}


\begin{figure}[!t]
\begin{center}
\plotfiddle{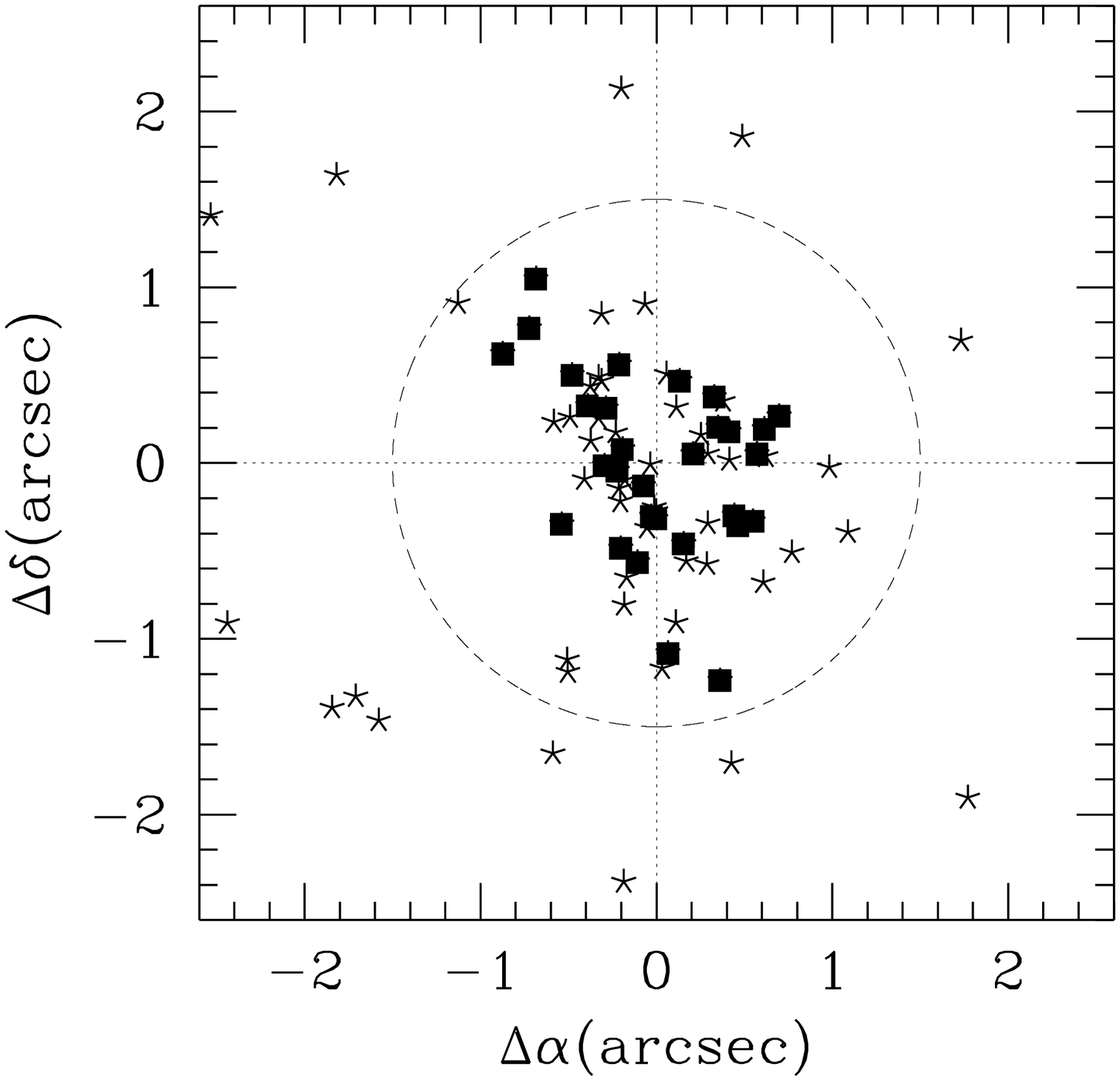}{1.8in}{0}{35}{35}{-230}{-100}
\plotfiddle{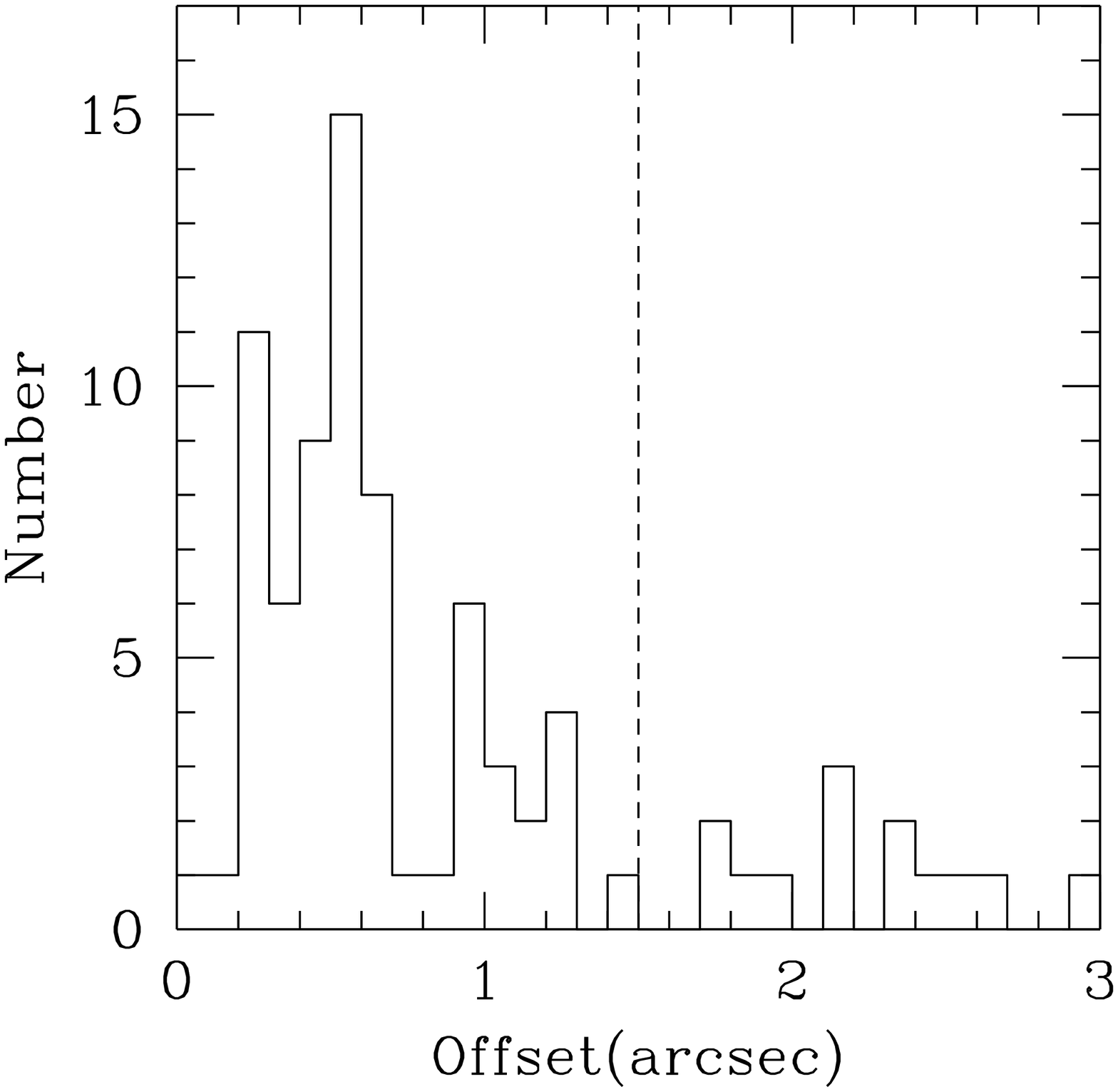}{0.0in}{0}{35}{35}{0}{-70}
\end{center}

\caption{Positional offsets between {\it Chandra} and $I$-band
identifications, after aligning the images.  Solid boxes in left panel
indicate the 30 bright X-ray sources ($>25$ counts in the full
0.5$-$7~keV image) with $17.5 < I < 22.5$ identifications which were
used to shift the {\it Chandra} map to the ground-based imaging.  Stars
indicate the nearest optical source to each X-ray source, regardless of
brightness.  We made host identifications using a 1\farcs5 matching
radius, indicated by the dashed line in both plots.}

\label{fig_offsets}
\end{figure}


\begin{figure}[!t]
\begin{center}
\plotfiddle{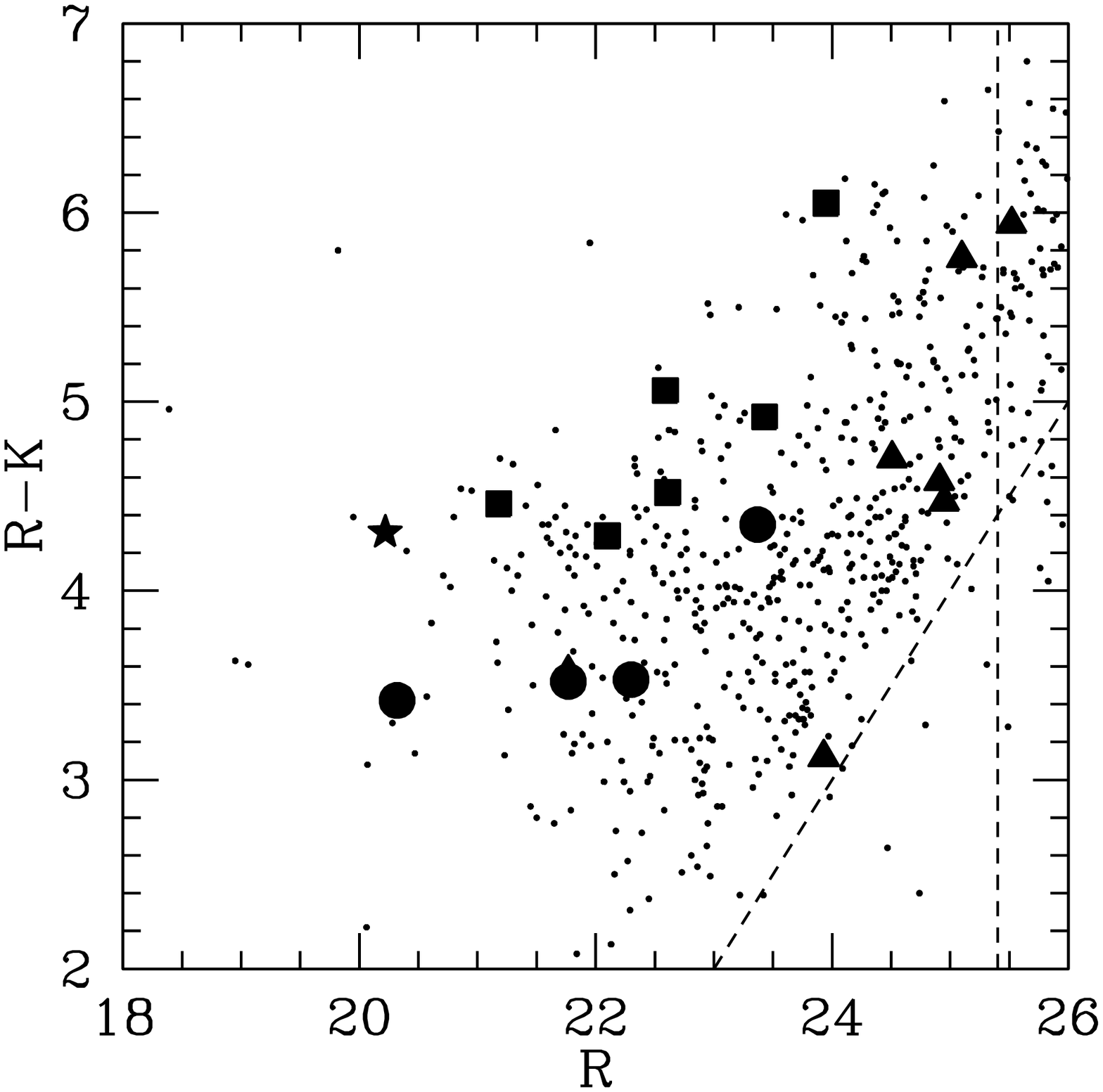}{1.8in}{0}{35}{35}{-230}{-100}
\plotfiddle{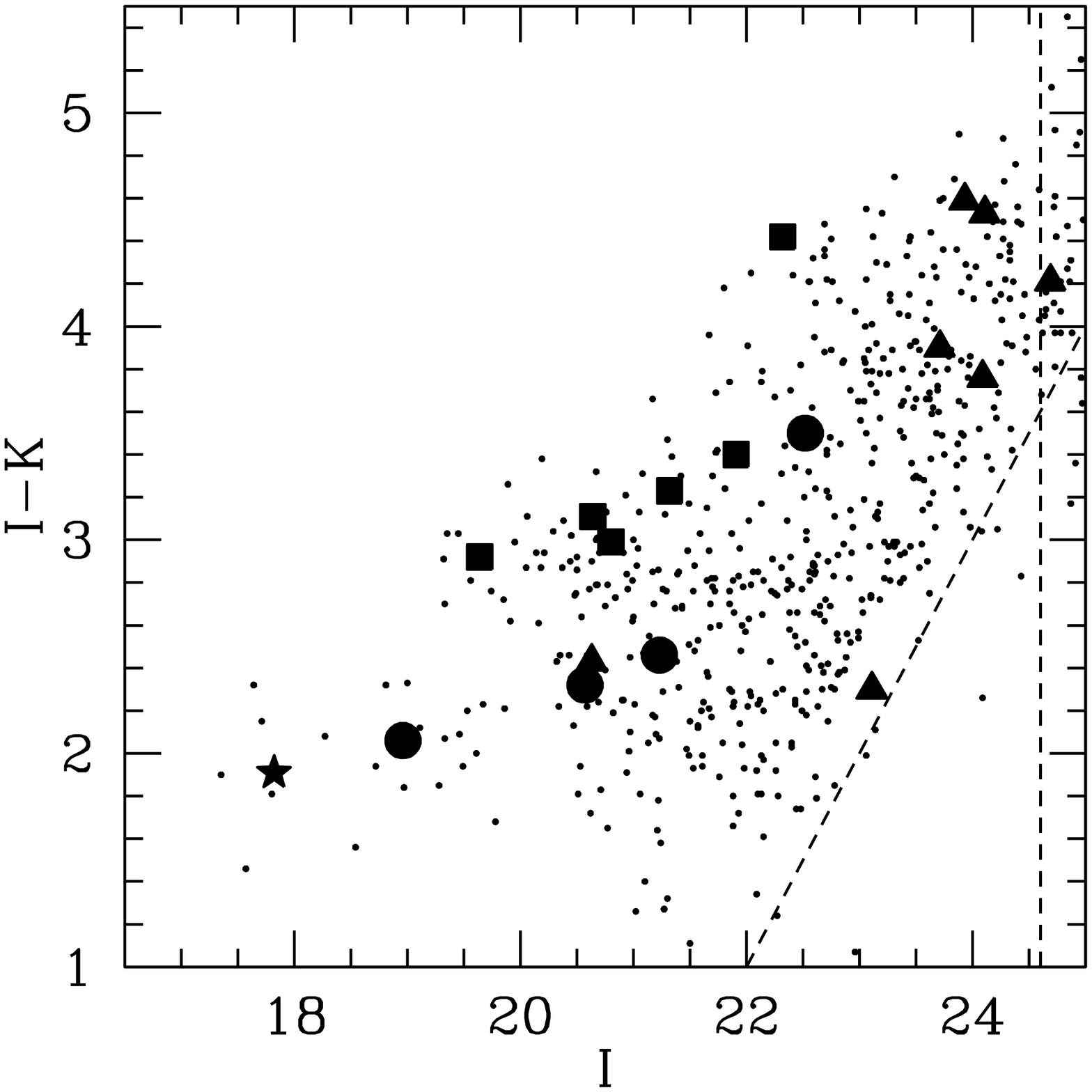}{0.0in}{0}{35}{35}{0}{-70}
\end{center}

\caption{Color-magnitude diagrams for sources in the central region of
the SPICES Lynx field.  Filled symbols represent X-ray sources from
this survey:  circles are spectroscopically identified with AGN,
squares are spectroscopically identified with apparently normal
galaxies, the star represents a Galactic M4 V dwarf, and triangles have
not been spectroscopically observed as yet.  Dots represent field
sources from the same region.  Dashed lines indicate the limits of our
imaging.  Note that the bluer sources tend to be associated with AGN,
while the redder sources tend to be associated with galaxies.}

\label{fig_IRcolor}
\end{figure}


\begin{figure}[!t]
\begin{center}
\end{center}

\caption{Images of the thirteen {\it Chandra} sources in the Lynx field
which were covered by our {\it HST}/WFPC2 F702W and F814W pointings.
Each panel is 12\arcsec\ square centered on the X-ray coordinates,
oriented with north to the top and east to the left.  Not all sources
are detected.  The exposure times are 2$-$10 orbits.  CXO52 is a Type~2
quasar at $z = 3.287$, discussed more extensively in Stern \etal
(2002).}

\label{fig_hst}
\end{figure}


\begin{figure}[!t]
\begin{center}
\plotfiddle{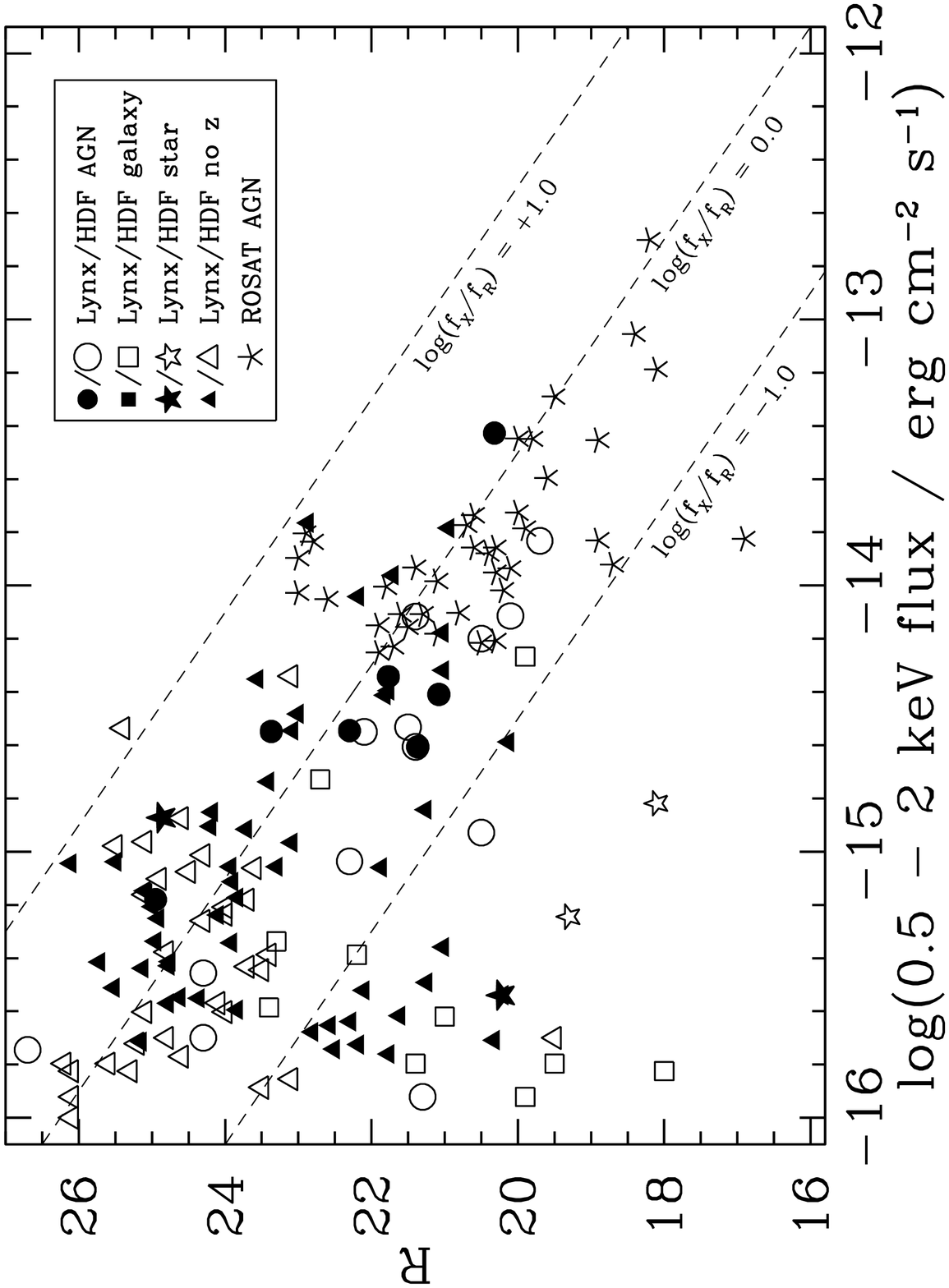}{2.9in}{-90}{40}{40}{-180}{260}
\plotfiddle{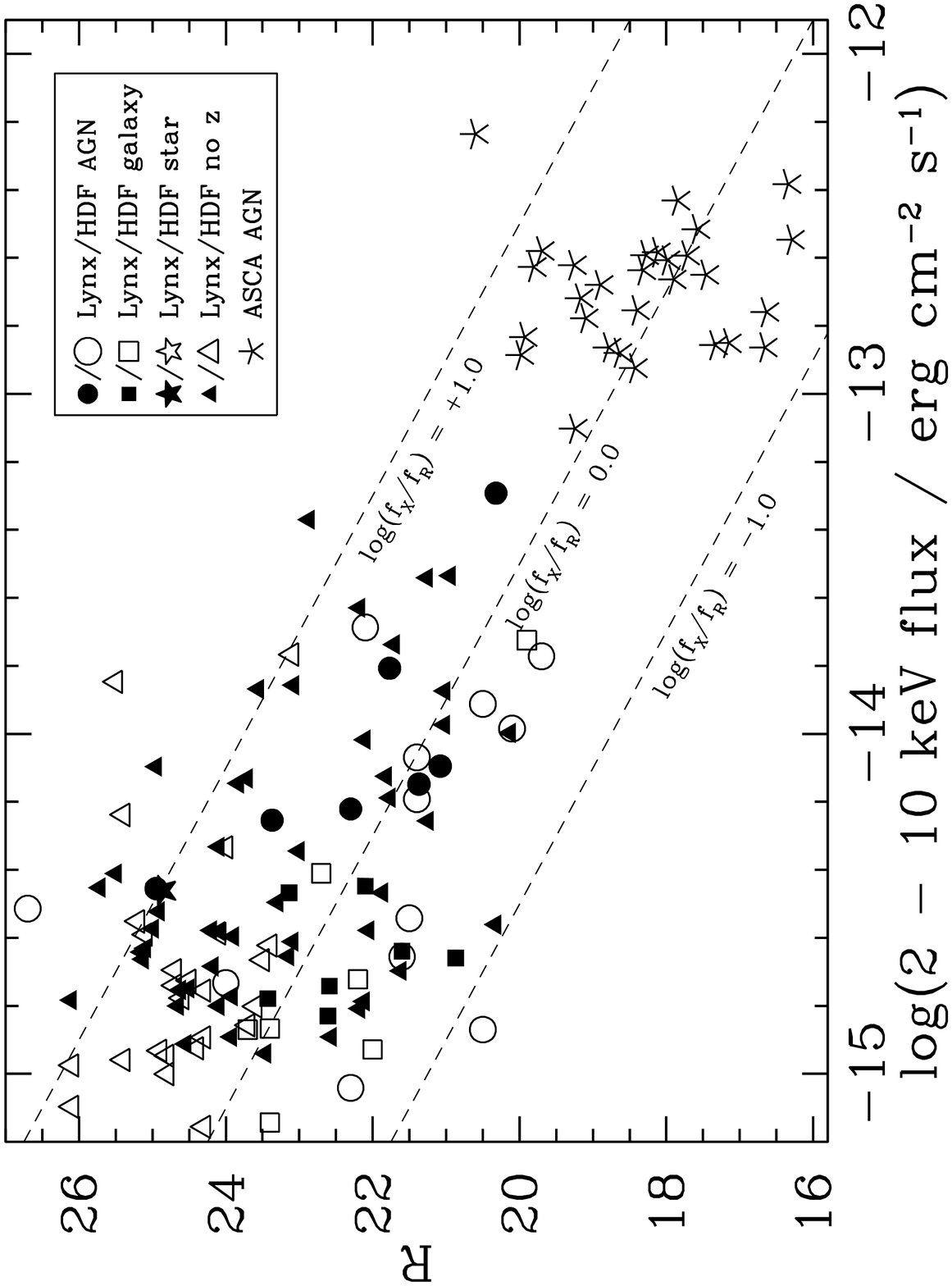}{2.9in}{-90}{40}{40}{-180}{240}
\end{center}

\caption{Optical $R$-band magnitude of X-ray selected sources, plotted
against soft ($0.5-2$~keV) and hard ($2-10$~keV) X-ray flux.  Filled
sources are from the {\it Chandra} Lynx field presented here.  Open
sources are from the 221.9~ks {\it Chandra} survey of the HDF-N
reported in Hornschemeier \etal (2001).  Asterisks at bright X-ray
fluxes show AGN from {\it ROSAT} for the soft X-ray band (Schmidt \etal
1998) and from {\it ASCA} for the hard X-ray band (Akiyama \etal
2000).  For the {\it Chandra} surveys, symbol shape corresponds to
spectroscopic classification:  circles are obvious AGN (\ie quasars and
sources with strong, high-ionization emission lines), squares are
apparently-normal galaxies, stars are Galactic dwarfs, and triangles
are spectroscopically unidentified sources.  We note that the
apparently-normal galaxies may harbour low-luminosity and/or
heavily-obscured AGN.  Slanted lines show location of constant
X-ray-to-optical flux ratio.}

\label{fig_Xopt}
\end{figure}


\begin{figure}[!t]
\begin{center}
\plotfiddle{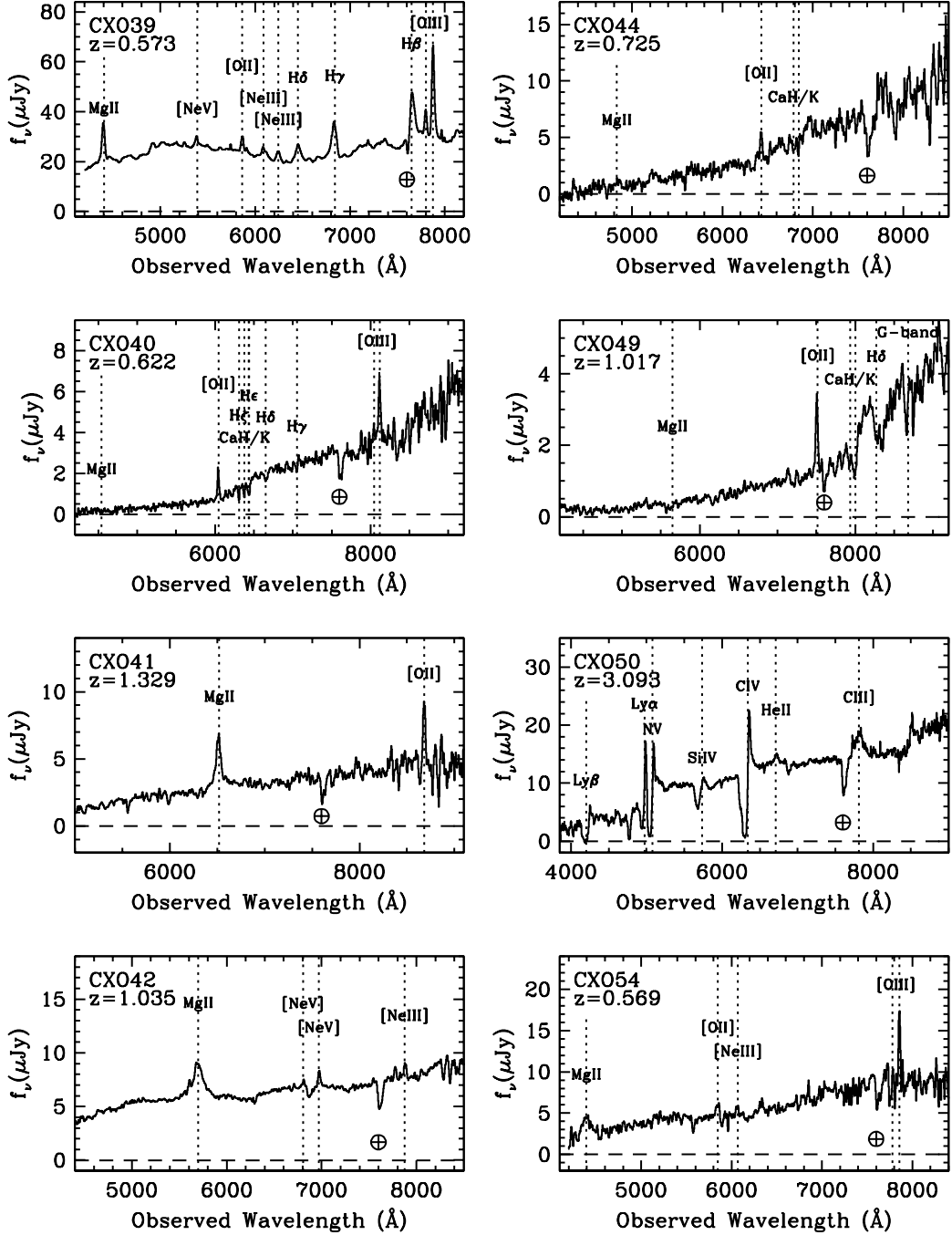}{6.5in}{0}{75}{75}{-240}{-50}
\end{center}

\caption{Spectra of {\it Chandra} sources in the Lynx field, obtained
with LRIS on the Keck telescopes.  Spectra were extracted using 1\farcs5
$\times$ 1\farcs5 apertures and have been smoothed with a boxcar filter,
typically of width 15 \AA.  Vertical dashed lines indicate the expected
wavelength of common spectroscopic features for the spectral class
determined in Table~3; not all are detected.}

\label{spectra}
\end{figure}



\begin{figure}[!t]
\begin{center}
\plotfiddle{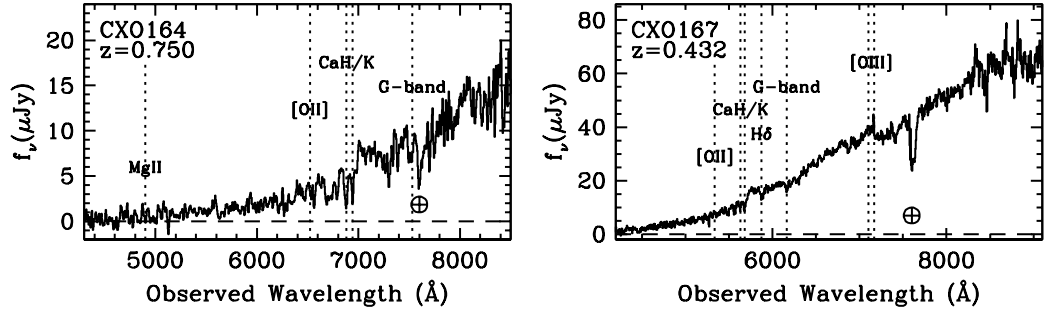}{6.5in}{0}{75}{75}{-240}{-50}
\plotfiddle{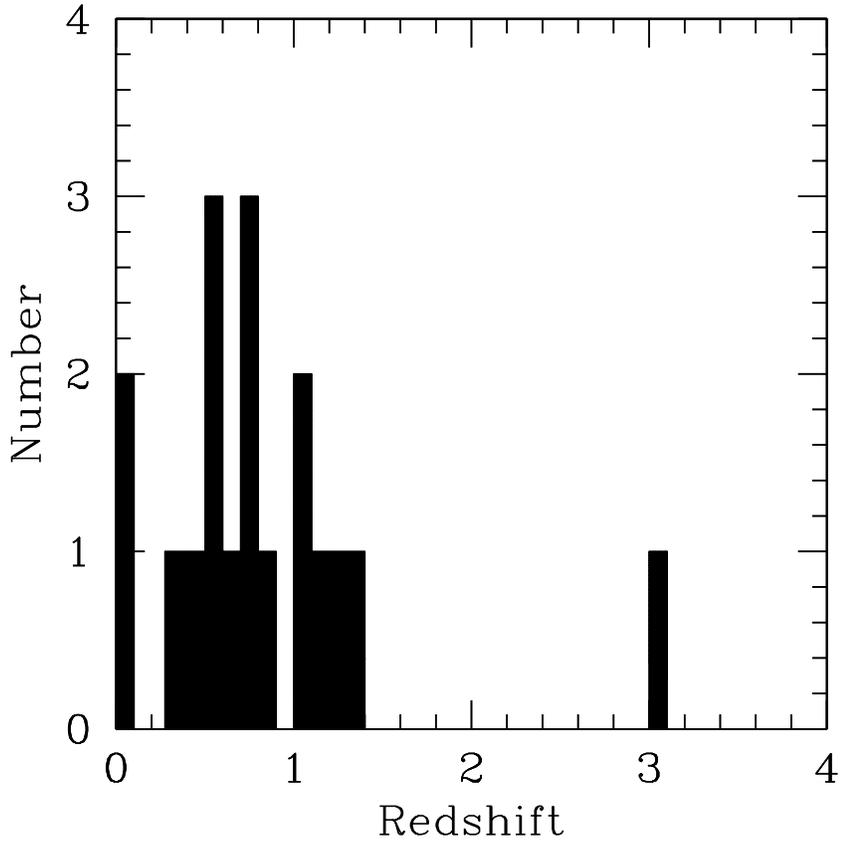}{0.0in}{0}{60}{60}{-200}{-80}
\end{center}

\caption{Redshift histogram of the 18 {\it Chandra} sources in the 
Lynx field for which we reveal spectroscopic information.}

\label{zhist}
\end{figure}


\begin{figure}[!t]
\begin{center}
\plotfiddle{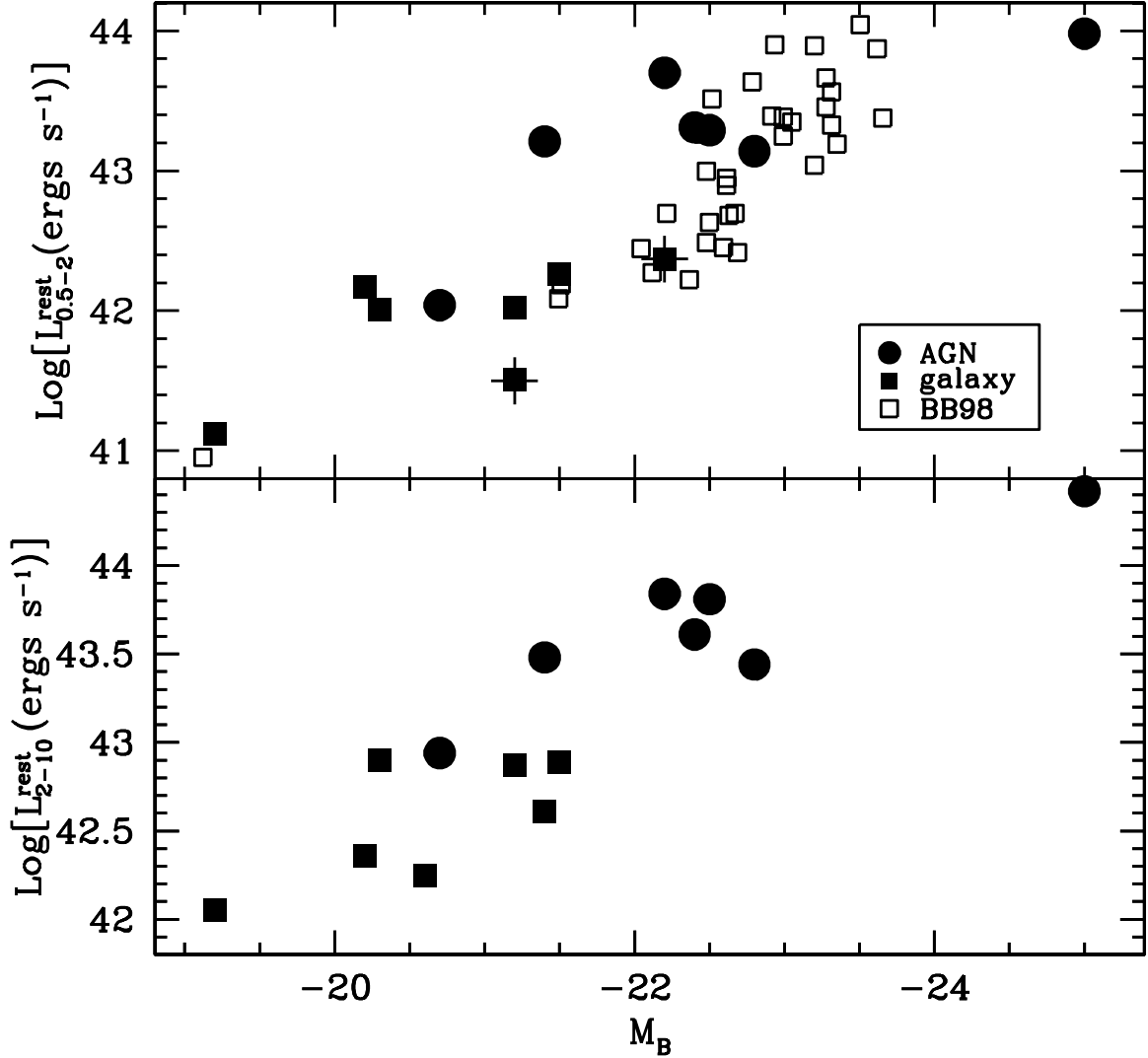}{4.6in}{0}{80}{80}{-250}{-140}
\end{center}

\caption{X-ray luminosity for the soft ($0.5-2$~keV) and hard
($2-10$~keV) bands plotted against rest-frame $B$-band luminosity for
the spectroscopic sample discussed here.  Symbol shape shows optical
spectral classification: solid circles correspond to sources with an
AGN obviously present, while solid squares represent optically-normal
galaxies.  The two early-type galaxies detected in the soft X-ray band
(CXO128 and CXO138) are marked with a plus sign in the upper panel.  In
the upper panel (soft X-ray band), we also include 34 bright early-type
galaxies observed by {\it ROSAT} from Brown \& Bregman (1998; BB98),
plotted as open squares.  See text for details.  All symbols are
plotted for an Einstein-de~Sitter universe with $H_0 = 50~ \kmsMpc$,
$\Omega_M = 1$, and $\Omega_\Lambda = 0$.  The figure is only slightly
changed for the dark energy cosmology favored by recent microwave
background and high-redshift supernovae experiments, $H_0 = 65~
\kmsMpc$, $\Omega_M = 0.35$, and $\Omega_\Lambda = 0.65$.}

\label{fig_Xlum}
\end{figure}


\begin{figure}[!t]
\begin{center}
\plotfiddle{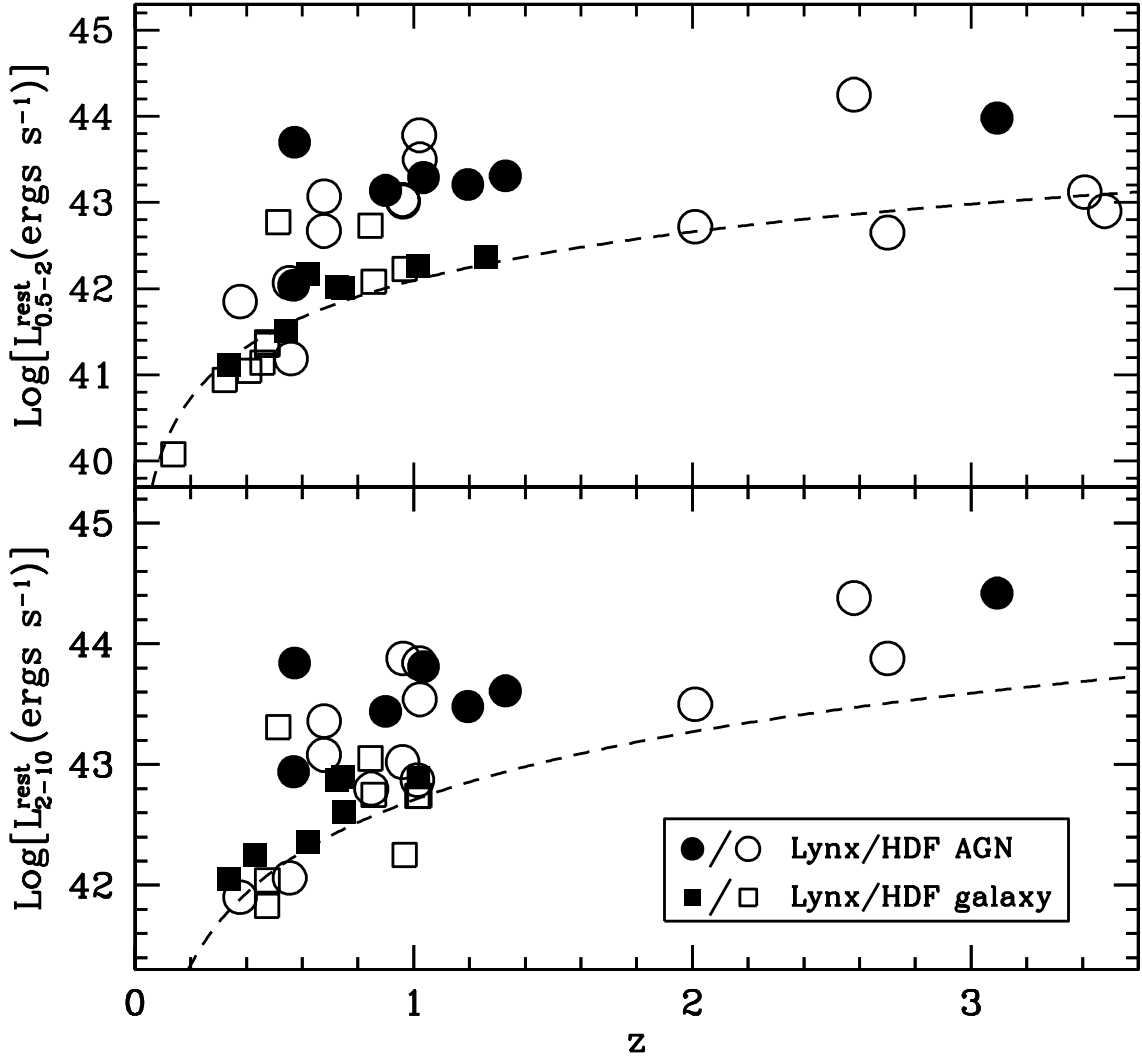}{4.6in}{0}{80}{80}{-250}{-140}
\end{center}

\caption{X-ray Luminosity for the soft ($0.5-2$~keV) and hard
($2-10$~keV) bands plotted against redshift for two samples of {\it
Chandra} sources.  Solid symbols are from this paper.  Open symbols are
from the HDF-N (Hornschemeier \etal 2001).  Luminosities for all
sources have been calculated for $\Omega = 1$, $\Lambda = 0$, and $H_0
= 50 \kmsMpc$, assuming an X-ray spectral index $\Gamma = 1.4$.  This
entailed recalculating the HDF-N points.  We also include the $z=2.010$
{\it Chandra} source in the HDF-N recently reported by Dawson \etal
(2001).  Symbol shape shows optical spectral classification:  circles
correspond to sources with an AGN obviously present, while squares
represent optically-normal galaxies.  Dashed lines show the flux limit
of our survey:  $S_{0.5-2} > 1.7 \times 10^{-16} \ergcm2s$ and
$S_{2-10} > 1.3 \times 10^{-15} \ergcm2s$.  The figure is only slightly
changed for the dark energy cosmology favored by recent microwave
background and high-redshift supernovae experiments, $H_0 = 65~
\kmsMpc$, $\Omega_M = 0.35$, and $\Omega_\Lambda = 0.65$.}

\label{fig_zlum}
\end{figure}


\begin{figure}[!t]
\begin{center}
\plotfiddle{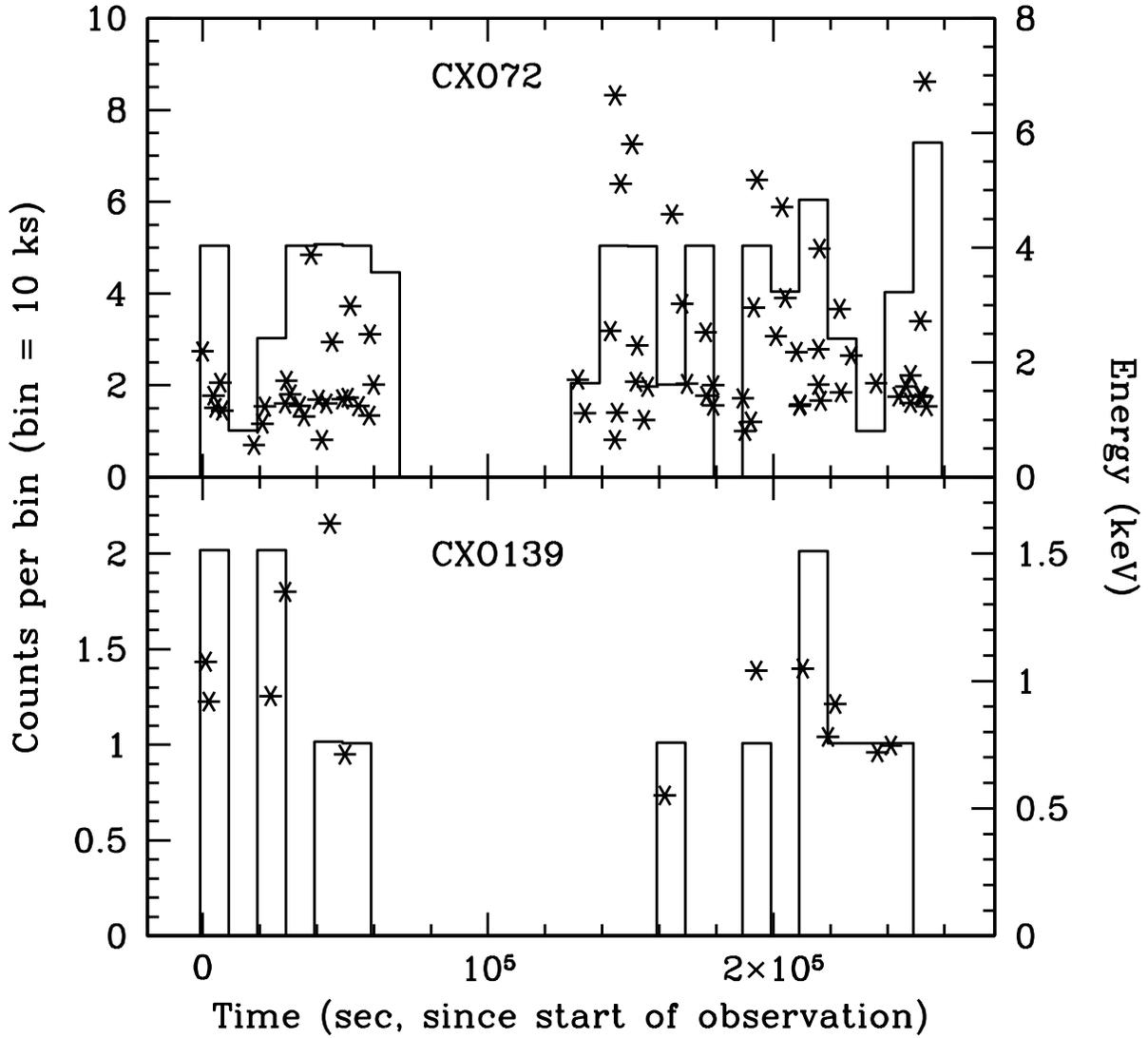}{4.6in}{0}{80}{80}{-250}{-140}
\end{center}

\caption{X-ray light curves for the two Galactic sources, the M7 dwarf
CXO72 and the M4 dwarf CXO139 (histogram; {\em left-hand scale}).  {\it
Chandra} observations of this field were conducted in two campaigns
(see \S 2.1); the gap between these campaigns is clearly evident,
centered at $10^6$~s after the start of observations.  The count rate
shows no evidence of variability or flaring for either source.
Asterisks mark the times and energies of the each photon detected ({\em
right-hand scale}).  The X-ray spectrum of CXO72 appears to steepen over
the course of our observations.}

\label{fig_lightcurve}
\end{figure}


\begin{figure}
\begin{center}
\plotfiddle{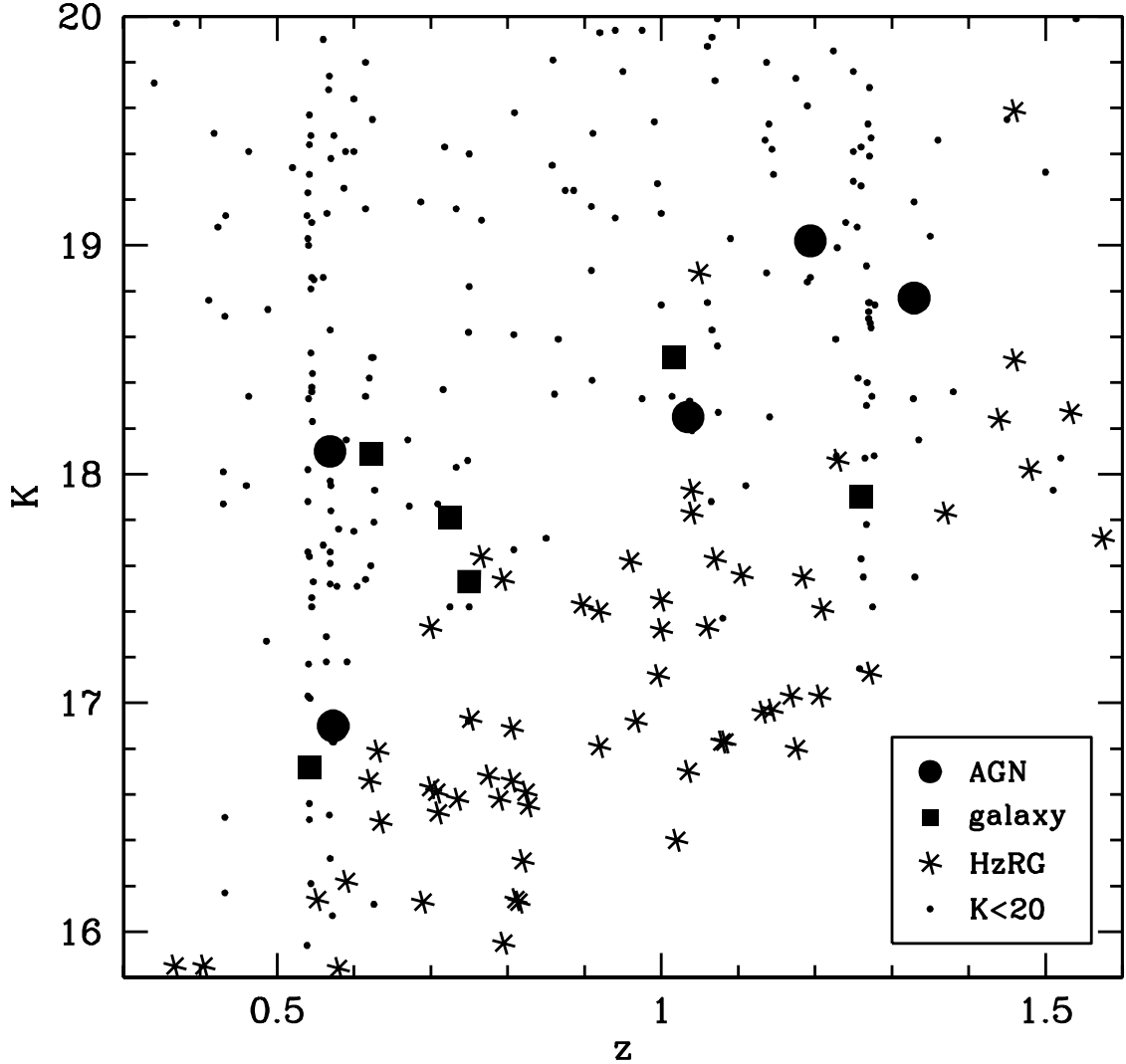}{4.6in}{0}{80}{80}{-250}{-140}
\end{center}

\caption{$K$-band brightness plotted against redshift.  Larger, solid
symbols refer to {\it Chandra} sources from this paper with symbol
shape referring to optical spectral classification:  circles refer to
obvious AGN while squares refer to optically-normal galaxies.
Asterisks are high-redshift radio galaxies (HzRGs) measured in 64~kpc
metric diameter apertures ($H_0 = 65 \kmsMpc$, $\Omega_m = 0.3$, and
$\Lambda = 0$) from De~Breuck \etal (2001).  Dots are galaxies from the
Lynx field of the $K$-selected ($K < 20$) SPICES survey, measured in
3\arcsec\ diameter apertures (Eisenhardt \etal 2001).  The differing
aperture sizes for the samples should produce systematic effects at
only a few tenths of a magnitude.  Note that the X-ray sources are
among the most luminous sources at 2$\mu$m for each redshift bin.  The
clusters at $z = 0.57$ and $z = 1.27$ are clearly evident in the field
galaxy redshift distribution.}

\label{fig_kz}
\end{figure}


\begin{thebibliography}{}

\bibitem[Akiyama {et~al.} 2000]{Akiyama:00}
Akiyama, M. {et~al.} 2000, \apj, 532, 700

\bibitem[Arnaud 1996]{Arnaud:96}
Arnaud, K.~A. 1996, in {\it Astronomical Data Analysis Software and Systems V},  ed. J.~G. Jacoby \& J.~Barnes, Vol. 101 (San Francisco: ASP Conference  Series), 17

\bibitem[Barger, Cowie, Mushotzky, \&  Richards 2001]{Barger:01}
Barger, A., Cowie, L.~L., Mushotzky, R.~F., \& Richards, E.~A. 2001, \aj,  submitted, astro-ph/0007175

\bibitem[Basri 1987]{Basri:87}
Basri, G. 1987, \apj, 316, 377

\bibitem[Bertin \& Arnouts 1996]{Bertin:96}
Bertin, E. \& Arnouts, S. 1996, \aap, 117, 393

\bibitem[Brown \& Bregman 1998]{Brown:98}
Brown, B. \& Bregman, J. 1998, \apj, 495, L75

\bibitem[Burgasser, Kirkpatrick, Reid, Liebert, Gizis,  \& Brown 2000]{Burgasser:00}
Burgasser, A.~J., Kirkpatrick, J.~D., Reid, I.~N., Liebert, J., Gizis, J.~E.,  \& Brown, M.~E. 2000, \aj, 120, 473

\bibitem[Cowie {et~al.} 2001]{Cowie:01}
Cowie, L. {et~al.} 2001, \apj, in press (astro-ph/0102306)

\bibitem[Dawson, Stern, Bunker, Spinrad, \&  Dey 2001]{Dawson:01}
Dawson, S., Stern, D., Bunker, A.~J., Spinrad, H., \& Dey, A. 2001, \aj,  accepted [astro-ph/0105043]

\bibitem[{De~Breuck}, {van~Breugel}, Stanford,  R\"ottgering, Miley, \& Stern 2001]{DeBreuck:01}
{De~Breuck}, C., {van~Breugel}, W., Stanford, S.~A., R\"ottgering, H., Miley,  G., \& Stern, D. 2001, \aj, submitted

\bibitem[{della~Ceca}, Castelli, Braito, Cagnoni,  \& Maccacaro 1999]{dellaCeca:99}
{della~Ceca}, R., Castelli, G., Braito, V., Cagnoni, I., \& Maccacaro, T. 1999,  \apj, 524, 674

\bibitem[Dey 1999]{Dey:99c}
Dey, A. 1999, in {\it The Most Disant Radio Galaxies}, ed. H.~R\"ottgering,  P.~N. Best, \& M.~D. Lehnert (Dordrecht: Kluwer), 19

\bibitem[Dunlop, Peacock, Spinrad, Dey, Jimenez, Stern,  \& Windhorst 1996]{Dunlop:96}
Dunlop, J.~S., Peacock, J.~A., Spinrad, H., Dey, A., Jimenez, R., Stern, D., \&  Windhorst, R.~A. 1996, \nat, 381, 581

\bibitem[Eisenhardt, Elston, Stanford, Stern, Wu,  Connolly, \& Spinrad 2001]{Eisenhardt:01}
Eisenhardt, P., Elston, R., Stanford, S.~A., Stern, D., Wu, K.~L., Connolly,  A., \& Spinrad, H. 2001, \aj, in preparation

\bibitem[Faber, Wegner, Burstein, Davies, Dressler,  Lynden-Bell, \& Terlevich 1989]{Faber:89}
Faber, S.~M., Wegner, G., Burstein, D., Davies, R.~L., Dressler, A.,  Lynden-Bell, D., \& Terlevich, R.~J. 1989, \apjs, 69, 763

\bibitem[Fanelli, O'Connell, Burstein, \&  Wu 1992]{Fanelli:92}
Fanelli, M.~N., O'Connell, R.~W., Burstein, D., \& Wu, C.~C. 1992, \apjs, 82,  197

\bibitem[Fiore, {La~Franca}, Giommi, Elvis, Matt,  Comastri, Molendi, \& Gioia 1999]{Fiore:99}
Fiore, F., {La~Franca}, F., Giommi, P., Elvis, M., Matt, G., Comastri, A.,  Molendi, S., \& Gioia, I. 1999, \mnras, 306, L55

\bibitem[Fiore, {La~Franca}, Vignali, Comastri, Matt,  Perola, Cappi, , Elvis, \& Nicastro 2000]{Fiore:00}
Fiore, F., {La~Franca}, F., Vignali, C., Comastri, A., Matt, G., Perola, G.~C.,  Cappi, M., , Elvis, {et al.}, 2000, New Astron., 5, 143

\bibitem[Forman, Jones, \& Tucker 1985]{Forman:85}
Forman, W., Jones, C., \& Tucker, W. 1985, \apj, 293, 102

\bibitem[Fowler, Gatley, Stuart, Joyce, \&  Probst 1988]{Fowler:88}
Fowler, A.~M., Gatley, I., Stuart, F., Joyce, R.~R., \& Probst, R.~G. 1988,  SPIE, 972, 107

\bibitem[Freeman, Kashyap, Rosner, \& Lamb]{Freeman:02}
Freeman, P.~E., Kashyap, V., Rosner, R., \& Lamb, D.Q. 2002, \apjs, in press 

\bibitem[Giacconi {et~al.} 2001]{Giacconi:01}
Giacconi, R. {et~al.} 2001, \apj, 551, 624

\bibitem[Giommi {et~al.} 1998]{Giommi:98}
Giommi, P. {et~al.} 1998, Nucl. Phys. B Proc. Suppl., 69, 591

\bibitem[Gizis, Monet, Reid, Kirkpatrick, Liebert, \&  Williams 2000]{Gizis:00}
Gizis, J.~E., Monet, D.~G., Reid, I.~N., Kirkpatrick, J.~D., Liebert, J., \&  Williams, R.~J. 2000, \aj, 120, 1085

\bibitem[Graham \& Dey 1996]{Graham:96}
Graham, J.~R. \& Dey, A. 1996, \apj, 471, 720

\bibitem[Hasinger, Burg, Giacconi, Hartner, Schmidt,  Tr\"umper, \& Zamorani 1993]{Hasinger:93}
Hasinger, G., Burg, R., Giacconi, R., Hartner, G., Schmidt, M., Tr\"umper, J.,  \& Zamorani, G. 1993, \aap, 275, 1

\bibitem[Hasinger, Burg, Giacconi, Schmidt, Tr\"umper,  \& Zamorani 1998]{Hasinger:98}
Hasinger, G., Burg, R., Giacconi, R., Schmidt, M., Tr\"umper, J., \& Zamorani,  G. 1998, \aap, 329, 482

\bibitem[Hasinger {et~al.} 2001]{Hasinger:01}
Hasinger, G. {et~al.} 2001, \aap, 365, 45

\bibitem[Hawarden, Leggett, Letawsky, Ballantyne, \&  Casali 2001]{Hawarden:01}
Hawarden, T.~G., Leggett, S.~K., Letawsky, M.~B., Ballantyne, D.~R., \& Casali,  M.~M. 2001, \mnras, 325, 563

\bibitem[Hawley, Gizis, \& Reid 1996]{Hawley:96}
Hawley, S.~L., Gizis, J.~E., \& Reid, I.~N. 1996, \aj, 112, 2799

\bibitem[Helfand \& Moran 2001]{Helfand:01}
Helfand, D.~J. \& Moran, E.~C. 2001, \apj, in press; astro-ph/0103484

\bibitem[Hogg 1999]{Hogg:99}
Hogg, D.~W. 1999, astro-ph/9905116

\bibitem[Holden, Stanford, Rosati, Tozzi, Eisenhardt, \&  Spinrad 2001]{Holden:01}
Holden, B., Stanford, S.~A., Rosati, P., Tozzi, P., Eisenhardt, P. R.~M., \&  Spinrad, H. 2001, \aj, in press (August)

\bibitem[Hornschemeier {et~al.} 2001]{Hornschemeier:01}
Hornschemeier, A.~E. {et~al.} 2001, \apj, in press; astro-ph/0101494

\bibitem[Ishisaki, Makishima, Takahashi, Ueda, Ogasaka,  \& Inoue 1999]{Ishisaki:99}
Ishisaki, Y., Makishima, K., Takahashi, T., Ueda, Y., Ogasaka, Y., \& Inoue, H.  1999, \apj, submitted

\bibitem[Kauffmann \& Charlot 1998]{Kauffmann:98}
Kauffmann, G. \& Charlot, S. 1998, \mnras, 294, 705

\bibitem[Kennefick, Djorgovski, \&  de~Calvalho 1995]{Kennefick:95}
Kennefick, J.~D., Djorgovski, S.~G., \& de~Calvalho, R.~R. 1995, \aj, 110, 2553

\bibitem[Kennicutt 1992]{Kennicutt:92}
Kennicutt, R. 1992, \apj, 388, 310

\bibitem[Kirkpatrick \& McCarthy 1994]{Kirkpatrick:94}
Kirkpatrick, J.~D. \& McCarthy, D.~W. 1994, \aj, 107, 333

\bibitem[Koyama, Hamaguchi, Ueno, Kobeyashi, \&  Feigelson 1996]{Koyama:96}
Koyama, K., Hamaguchi, K., Ueno, S., Kobeyashi, N., \& Feigelson, E. 1996,  \pasj, 48, L87

\bibitem[Kraft 1967]{Kraft:67}
Kraft, R.~P. 1967, \apj, 150, 551

\bibitem[Landolt 1992]{Landolt:92}
Landolt, A.~U. 1992, \aj, 104, 340

\bibitem[Leitherer, Carmelle, \&  Heckman 1995]{Leitherer:95}
Leitherer, C., Carmelle, R., \& Heckman, T.~M. 1995, \apjs, 99, 173

\bibitem[Lilly, Fevre, Hammer, \& Crampton 1996]{Lilly:96}
Lilly, S.~J., Fevre, O.~L., Hammer, F., \& Crampton, D. 1996, \apj, 460, L1

\bibitem[Marshall {et~al.} 1980]{Marshall:80}
Marshall, F. {et~al.} 1980, \apj, 235, 4

\bibitem[Massey \& Gronwall 1990]{Massey:90}
Massey, P. \& Gronwall, C. 1990, \apj, 358, 344

\bibitem[Matsumoto, Koyama, Awiki, Tsuru, Loewenstein,  \& Matsushita 1997]{Matsumoto:97}
Matsumoto, H., Koyama, K., Awiki, H., Tsuru, T., Loewenstein, M., \&  Matsushita, K. 1997, \apj, 482, 133

\bibitem[McCarthy 1993]{McCarthy:93}
McCarthy, P.~J. 1993, \araa, 31, 639

\bibitem[Mushotzky, Cowie, Barger, \&  Arnaud 2000]{Mushotzky:00}
Mushotzky, R.~F., Cowie, L.~L., Barger, A.~J., \& Arnaud, K.~A. 2000, \nat,  404, 459

\bibitem[Nakanishi, Akiyama, Ohta, \& Yamada 2000]{Nakanishi:00}
Nakanishi, K., Akiyama, M., Ohta, K., \& Yamada, T. 2000, \apj, 534, 587

\bibitem[Norman {et~al.} 2001]{Norman:01}
Norman, C. {et~al.} 2001, \apj, submitted, astro-ph/0103198

\bibitem[Ohta, Yamada, Nakanishi, Ogasaka, Kii, \& Hayashida 1996]{Ohta:96}
Ohta, K., Yamada, T., Nakanishi, K., Ogasaka, Y., Kii, T., \& Hayashida, K. 1996, \apj, 458, L57

\bibitem[Oke 1974]{Oke:74}
Oke, J.~B. 1974, \apjs, 27, 21

\bibitem[Oke {et~al.} 1995]{Oke:95}
Oke, J.~B. {et~al.} 1995, \pasp, 107, 375

\bibitem[Pallavicini, Peres, Serio, Vaiana, Golub,  \& Rosner 1987]{Pallavicini:87}
Pallavicini, R., Peres, G., Serio, S., Vaiana, G., Golub, L., \& Rosner, R.  1987, \apj, 248, 279

\bibitem[Parker 1955]{Parker:55}
Parker, E.~N. 1955, \apj, 122, 293

\bibitem[Richstone \& Schmidt 1980]{Richstone:80}
Richstone, D.~O. \& Schmidt, M. 1980, \apj, 235, 361

\bibitem[Riess {et~al.} 2001]{Riess:01}
Riess, A.~G. {et~al.} 2001, \apj, in press; astro-ph/010455

\bibitem[Rosati, {della~Ceca}, Burg, Norman, \&  Giacconi 1995]{Rosati:95}
Rosati, P., {della~Ceca}, R., Burg, R., Norman, C., \& Giacconi, R. 1995, \apj,  445, L11

\bibitem[Rosati, {della~Ceca}, Norman, \&  Giacconi 1998]{Rosati:98}
Rosati, P., {della~Ceca}, R., Norman, C., \& Giacconi, R. 1998, \apj, 492, L21

\bibitem[Rosati, Stanford, Eisenhardt, Elston, Spinrad,  Stern, \& Dey 1999]{Rosati:99}
Rosati, P., Stanford, S.~A., Eisenhardt, P.~R., Elston, R., Spinrad, H., Stern,  D., \& Dey, A. 1999, \aj, 118, 76

\bibitem[Rutledge, Basri, Mart\'in, \&  Bildsten 2000]{Rutledge:00}
Rutledge, R., Basri, G., Mart\'in, E.~L., \& Bildsten, L. 2000, \apj, 538, L141

\bibitem[Sarazin, Irwin, \& Bregman 2000]{Sarazin:00}
Sarazin, C., Irwin, J., \& Bregman, J. 2000, \apj, 544, 101

\bibitem[Schmidt, Hasinger, Gunn, Schneider, Burg,  Giacconi, Lehmann, MacKenty, Tr\"umper, \& orani 1998]{Schmidt:98}
Schmidt, M., Hasinger, G., Gunn, J., Schneider, D., Burg, R., Giacconi, R.,  Lehmann, I., MacKenty, J., {et al.}, 1998, \aap, 329,  495

\bibitem[Schneider, van Gorkom, Schmidt, \&  Gunn 1992]{Schneider:92}
Schneider, D.~P., van Gorkom, J.~H., Schmidt, M., \& Gunn, J.~E. 1992, \aj,  103, 1451

\bibitem[Schreier {et~al.} 2001]{Schreier:01}
Schreier, E.~J. {et~al.} 2001, \apj, in press, astro-ph/0106248

\bibitem[Spinrad, Dey, Stern, Peacock, Dunlop, Jimenez,  \& Windhorst 1997]{Spinrad:97}
Spinrad, H., Dey, A., Stern, D., Peacock, J.~A., Dunlop, J., Jimenez, R., \&  Windhorst, R.~A. 1997, \apj, 484, 581

\bibitem[Stanford, Elston, Eisenhardt, Spinrad, Stern,  \& Dey 1997]{Stanford:97}
Stanford, S.~A., Elston, R., Eisenhardt, P.~R.~M., Spinrad, H., Stern, D., \&  Dey, A. 1997, \aj, 114, 2232

\bibitem[Stanford, Rosati, Holden, Tozzi, Eisenhardt,  \& Spinrad 2001]{Stanford:01}
Stanford, S.~A., Holden, B., Rosati, P., Tozzi, P., Borgani, S., Eisenhardt, P.~R.~M., \&  Spinrad, H. 2001, \apj, 552, 504

\bibitem[Stern, Bunker, Spinrad, \& Dey 2000]{Stern:00d}
Stern, D., Bunker, A.~J., Spinrad, H., \& Dey, A. 2000, \apj, 537, 73

\bibitem[Stern, Dey, Spinrad, Maxfield, Dickinson,  Schlegel, \& Gonz\'alez 1999]{Stern:99a}
Stern, D., Dey, A., Spinrad, H., Maxfield, L.~M., Dickinson, M.~E., Schlegel,  D., \& Gonz\'alez, R.~A. 1999, \aj, 117, 1122

\bibitem[Stern {et~al.} 2002]{Stern:02}
Stern, D. {et~al.} 2002, \apj, submitted

\bibitem[Tozzi {et~al.} 2001]{Tozzi:01}
Tozzi, P. {et~al.} 2001, \apj, in press (astro-ph/0103014)

\bibitem[Trauger {et~al.} 1994]{Trauger:94}
Trauger, S. {et~al.} 1994, \apj, 435, 3

\bibitem[Ueda {et~al.} 1999]{Ueda:99}
Ueda, Y. {et~al.} 1999, \apj, 518, 656

\bibitem[van Dokkum, Stanford, Holden, Eisenhardt,  Dickinson, \& Elston 2001]{vanDokkum:01}
van Dokkum, P.~G., Stanford, S.~A., Holden, B.~P., Eisenhardt, P.~R.,  Dickinson, M., \& Elston, R. 2001, \apj, in press

\bibitem[Vecchi, Molendi, Guainazzi, Fiore, \&  Parmar 1999]{Vecchi:99}
Vecchi, A., Molendi, S., Guainazzi, M., Fiore, F., \& Parmar, A.~N. 1999, \aap,  349, 73

\bibitem[Weisskopf, {O'dell}, \&  {van~Speybroeck} 1996]{Weisskopf:96}
Weisskopf, M.~C., {O'dell}, S.~L., \& {van~Speybroeck}, L.~P. 1996, SPIE, 2805,  2

\end{thebibliography}
\end{document}